\documentclass[aps,physrev,preprint,a4paper,nofootinbib,superscriptaddress,notitlepage]{revtex4-2}
\usepackage[latin1]{inputenc}
\usepackage[T1]{fontenc}
\usepackage{amsmath}
\usepackage{amsfonts}
\usepackage{amssymb}
\usepackage{graphicx,color}
\usepackage[margin=2cm]{geometry}
\usepackage{bm}
\usepackage{hyperref}
\usepackage{graphicx}
\usepackage{marginnote}
\usepackage{siunitx}
\usepackage{subcaption}
\usepackage{orcidlink}

\DeclareMathOperator{\sech}{sech}
\newenvironment{rcases}{\left.\begin{aligned}} {\end{aligned}\right\rbrace}

\begin{document}

\newcommand{\revision}[1]{{#1}}

\newcommand{\Da}{\text{Da}}
\newcommand{\Hankel}[1]{\mathcal{H}_0\left[ #1 \right]}
\newcommand{\Hankeln}[1]{\mathcal{H}_n\left[ #1 \right]}
\newcommand{\zok}{\quad\textrm{(0ok)}}
\newcommand{\fok}{\quad\textrm{(1ok)}}

\title{Colloidal bubble propulsion mediated through viscous flows}
\date{\today}
\author{Alexander Chamolly \orcidlink{0000-0002-2383-9314}}
\email{alexander.chamolly@pasteur.fr}
\affiliation{Institut Pasteur, Université Paris Cité, CNRS UMR3738, Developmental and Stem Cell Biology Department, F-75015 Paris, France}
\affiliation{Laboratoire de Physique de l'École normale supérieure, ENS, Université PSL, CNRS, Sorbonne Université, Université Paris Cité, F-75005 Paris, France}
\author{S\'ebastien Michelin \orcidlink{0000-0002-9037-7498}}
\email{sebastien.michelin@polytechnique.edu}
\affiliation{LadHyX, CNRS -- Ecole Polytechnique, Institut Polytechnique de Paris, F-91128 Palaiseau Cedex, France}
\author{Eric Lauga  \orcidlink{0000-0002-8916-2545}}
\email{e.lauga@damtp.cam.ac.uk}
\affiliation{Department of Applied Mathematics and Theoretical Physics, University of Cambridge, Wilberforce Road, CB3 0WA, Cambridge, UK}

\begin{abstract}
	Bubble-propelled catalytic colloids stand out as a uniquely efficient design for artificial controllable micromachines, but so far lack a  general theoretical framework that explains the physics of their propulsion. 
	Here we develop a combined diffusive and hydrodynamic theory of bubble growth near a spherical catalytic colloid, that allows us to explain the underlying mechanism and the influence of environmental and material parameters. We identify two dimensionless groups, related to colloidal activity and the background fluid, that govern a saddle-node bifurcation of the bubble growth dynamics, and calculate the generated flows analytically for both slip and no slip boundary conditions on the bubble. We finish with a discussion of the assumptions and predictions of our model in the context of existing experimental results, and conclude that some of the observed behaviour, notably the ratchet-like gait, stems from peculiarities of the experimental setup rather than fundamental physics of the propulsive mechanism. 
\end{abstract}

\maketitle
\def\v{\vspace{2cm}}
\section{Introduction}
	
Since their introduction in 2004~\cite{paxton04}, catalytic micromotors have been subject of intense theoretical and experimental investigation~\cite{elgeti2015physics,illien2017fuelled,moran2017phoretic}.  This interest is driven in part  by their potential in a wide range of applications, including biomedical uses such as drug delivery~\cite{nelson2010microrobots} or stem cell transplantation~\cite{jeon2019magnetically}. Initially,  investigations focused mostly  on theoretical~\cite{golestanian2007designing,chamolly2019stochastic,chamolly2020irreversible} and material-science aspects~\cite{tottori2012magnetic,walther2012soft}. Recently, practical uses of these active motors have  been achieved~\cite{li2018development}, facilitated in part by the introduction of bio-compatible designs~\cite{soto2021smart}. In many cases, these microrobots are immersed in a fluid medium, and  exploit (or are subject to) the particular physics of hydrodynamics at such small scales. To this end, several physical mechanisms to generate propulsion have been proposed, including external actuation with magnetic fields~\cite{ghosh2009controlled} and ultrasound~\cite{mou2015single}, as well as autonomous swimming through the exploitation of phoretic effects~\cite{michelin2014phoretic,moran2011electrokinetic}, and bubble propulsion~\cite{gibbs2009autonomously,manjare2012bubble,manjare2013bubble,gallino2018physics}. While phoretic propulsion relies on a fluid flow that is driven by an osmotic imbalance of ions produced on the swimmer's surface in an asymmetric fashion, bubble propulsion is a variant of this mechanism in which the produced fluxes are so large that locally a supersaturation is achieved, which makes the nucleation of a gas bubble energetically favourable. Rather than the weak coupling between chemical gradients and fluid mechanics \cite{michelin2014phoretic}, it is then a direct mechanical coupling with the bubble growth dynamics that propels the swimmer. As a result, the achieved propulsion speeds are much higher, on the order of $\si{\milli\metre\per\second}$, rather than $\si{\micro\meter\per\second}$ \cite{moran2017phoretic}. 
  
Perhaps due to these empirical observation of fast motion, theoretical work investigating bubble-propelled swimmers has  focused for the most part on inertial models, such as momentum transfer from the detachment of nanobubbles~\cite{gibbs2009autonomously}, and inertial bubble collapse~\cite{manjare2012bubble}. Attempts have been made to rationalise bubble propulsion dynamics by comparing with the phenomenology of bubble growth as simulated with Lattice-Boltzmann methods and investigated experimentally~\cite{yang2001numerical,zeng1993unified}, but since these had been performed in a regime where the Reynolds number is at least of order unity, it remains doubtful whether they capture the correct physics. Indeed, in order to compensate for the discrepancy of scales, any quantitative comparisons have required the adjustment of empirical coefficients for the forces that growing bubbles exert on boundaries, normally of order unity, to several orders of magnitude larger~\cite{manjare2012bubble,manjare2013bubble}. Other models for bubble-propulsion have given significant attention to surface tension, mass and momentum conservation, but did not incorporate fluid mechanical modelling of the interaction between bubble and colloid~\cite{fomin2013propulsion,li2014hydrodynamics}. 
	
While flow physics were incorporated  numerically in the specific case of conical microrockets~\cite{wang2017viscosity,wang2018dynamic,gallino2018physics}, no theoretical work has been devoted towards gaining a fundamental understanding of the propulsion mechanism for the classical  design of a spherical colloidal catalyst powered by a collection of nucleated bubbles. Modelling needs to include (1) the  physics of bubble growth itself, which due to the small scales exhibit peculiarities that are different from macroscopic dynamics~\cite{lohse2015surface,rodriguez2015generation} and (2)  the transport within the realm of inertia-less hydrodynamics, in which propulsion is not achieved by momentum exchange with the surrounding fluid, but through non-reciprocal changes in the geometry~\cite{happel2012low,lauga2009hydrodynamics}, here due to bubble growth.
	
	\begin{figure}[t]
		\centering
		\includegraphics[width=0.6\textwidth]{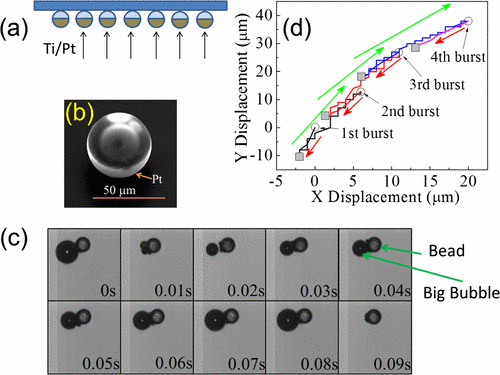}
		\caption{Schematic of the experimental setup by Ref.~\cite{manjare2012bubble}. (a) Manufacture of beads, (b) Scanning electron microscope image of the swimmer. (c) Snapshots of a microbead showing the bubble growth and burst processes and the bead motion behavior. (d) The trajectory of the bead. The red arrows denote the direction of the trajectories of the bead after bubble burst, and the green arrows represent the direction of the trajectories of the bead during bubble growth. Adapted from Manjare, Yang and  Zhao (2012) {\it Phys.~Rev.~Lett.}, {\bf 109}, 128305~\cite{manjare2012bubble} with permission.}\label{fig:experiments}
	\end{figure}
	 	
The canonical experimental setup is illustrated Fig.~\ref{fig:experiments}, adapted from Ref.~\cite{manjare2012bubble}, including  (a) the production of the microswimmer, (b) an image of the swimmer \revision{featuring the smooth Pt-covered reactive surface}, (c) a time-lapse of the bubble growth dynamics, and (d) the resulting trajectory. Bubbles are nucleated on the surface of the spherical colloid and propulsion occurs in a two-step ratchet-like periodic motion where slow pushing due to diffusive bubble growth is followed by fast retraction due to sudden bubble disappearance. In this work, we present a theoretical study to address this two-step motion. We first \revision{derive} a coupled model of diffusive bubble growth and inertialess hydrodynamics in terms of experimentally relevant parameters in section \ref{sec:diff}. \revision{Here we also include a discussion of the model assumptions and their experimental relevance}. We then solve this model, first for the instantaneous diffusive growth dynamics and instantaneous coupling with fluid dynamics, followed by their combined time-dependent evolution in section \ref{sec:results}. We comment briefly on the conservation of momentum during inertial bubble collapse in \ref{sec:inertia} and make a prediction for a new type of bubble-propelled swimmer that could be experimentally realised. We conclude with a summary in section \ref{sec:discussion}.

\section{Mathematical modelling of bubble growth and propulsion}\label{sec:diff}
	\subsection{Motivation}
	
	We begin by building some intuition for the physics of the growth process based on the typical length and time scales involved. The colloidal particle and the bubble are small, with a relevant length scale $L\sim\mathcal{O}(10)\si{\micro\metre}$, while the experimentally observed time scale of the bubble growth cycle is about $t\sim0.1\si{\second}$~\cite{manjare2012bubble}. Assuming a surface tension $\gamma\approx\SI{7.2e-2}{\newton\per\metre}$ as for a clear air-water interface, a mass density for the fluid of $\rho_l\approx \SI{1e3}{\kilogram\per\metre\cubed}$ and a dynamic fluid viscosity of $\mu \approx \SI{1e-3}{\pascal\second}$, we can compute the dimensionless Reynolds and Capillary numbers as
	\begin{align}
		\text{Re} \sim\frac{\rho_l L^2}{\mu t} \sim 10^{-3},\quad \text{Ca} \sim\frac{\mu L}{\gamma t} \sim 10^{-6}.
	\end{align}
	From this we deduce that inertial effects and the contribution of the dynamic pressure to the total pressure in the bubble are both negligible. This implies that the bubble does remain spherical throughout its growth (as is observed in experiments). We can thus write at all time that the bubble pressure $p_\text{bubble}$ is given by the Laplace law \cite{rodriguez2015generation}
\begin{align}\label{eq:Lapla}
	p_\text{bubble}= p^\infty + \frac{2\gamma}{R_B},
\end{align}
where $p^\infty$ is the hydrostatic background pressure and $R_B$ is the radius of the bubble. This is the inertia-less version of the general Rayleigh-Plesset equation that is canonically used for the description of bubble growth. In this limit the bubble growth rate $\dot{R}_B$ does not enter the equation. As a result, the bubble size varies quasi-steadily with the bubble pressure. 
	
	The pressure in the bubble $p_\text{bubble}$  is in turn related to the (molar) gas density $\rho_g$ via the ideal gas law as
	\begin{align}\label{eq:idealgaslaw}
		p_\text{bubble}=\rho_g\mathcal{R}T,
	\end{align}
	where $\mathcal{R}$ is the universal gas constant and $T$ is the absolute temperature, which we assume to be uniform and constant in time. To determine the density \revision{$\rho_g(t)$ as a function of time}, we thus need to solve for the molar concentration field $c$ of dissolved gas surrounding the system, and then compute the mass flux into the bubble.

	The two fundamental physical processes at work in the system are therefore (i) the production and transport of gas into the bubble, leading to its growth, and (ii) hydrodynamic interactions between the bubble and the colloidal particle that convert bubble growth into propulsion. In general these two problems are coupled mathematically, since the flux-dependent bubble growth rate determines the fluid flows, and these in turn can affect the transport of dissolved gas through advection. 
	However, in the limit where gas transport is dominated by diffusion  the latter effect is negligible, so there is only a one-way coupling:  a quasi-steady problem may be solved first to find the instantaneous bubble growth rate, and its solution can then be fed into the equations of fluid mechanics to determine the propulsion velocity. In our discussion of parameter estimates in section \ref{sec:paramestimates}, we show that this condition is adequately satisfied for the canonical decomposition of hydrogen peroxide in which the surface of the colloid acts as a catalyst and the produced gas is oxygen ($2\, \rm{H_2O_2} \to \rm{O_2}+2\, \rm{H_2O}$) as used in many experimental studies~\cite{manjare2012bubble,elgeti2015physics,moran2017phoretic}.
	
	\revision{
	\subsection{Assumptions}\label{sec:assumptions}
	
	In the course of developing and analysing this model we make a number of assumptions, which we briefly summarise here.
	
	Since the most commonly used experimental swimmers feature a Pt-coated surface acting as a catalyst for the chemical reaction \cite{moran2017phoretic,manjare2012bubble}, we assume that the reactive surface is smooth and not porous. While examples of nanoporous colloids exist \cite{zhai2019precise,kaewsaneha2013janus}, these have not to our knowledge been shown to self-propel, or designed with the goal of achieving self-propulsion. While the raw silica beads used in the production of phoretic swimmers do appear to exhibit significant surface roughness in SEM images, this roughness is not visible after Pt-coating (see Fig.~\ref{fig:experiments}(b)). It is conceivable that the nature of the coating process tends to smooth out such irregularities, and we hence ignore them in this work.
	
	Furthermore, previous theoretical models of (phoretic) swimming have occasionally included explicitly a thin ionic interaction layer of width $\sim\SI{1}{\nano\meter}$ above the reactive surface \cite{michelin2014phoretic,sabass2012dynamics,sharifi2013diffusiophoretic}. Since this layer is negligibly thin compared to the colloid and bubble radii, we also ignore it here. While it may be important locally at the contact point, this effect would amount only to a higher-order correction to the global fluxes computed in this study. For the reaction itself, we assume in the main text that a uniform constant flux of solute is produced on the colloid, and additionally discuss quantitatively  the higher-order correction due to considering only a constant reaction rate in Appendix \ref{sec:1ok}.
	
	For mathematical simplicity, we further assume that there is only a single bubble present in the problem, and we consider only the dynamics post-nucleation. We note indeed that experimentally, while there are occasionally multiple bubbles forming, there is usually one that emerges to dominate in size~\cite{manjare2012bubble}. At early stages, this process is likely driven by Ostwald ripening~\cite{inoue2022ostwald}. At later stages, local perturbations of the concentration field due to the presence of small bubbles may lead to short-term deviations on the time scale of dissolution. Modelling these additional bubbles mathematically is challenging, and it is unclear whether they enhance or inhibit propulsion. Their contribution would thus deserve an analysis that is beyond the scope of the present paper. Additionally, we ignore any variation in surface tension on the bubble that may arise, for instance, from a spatial heterogeneities in the chemical reaction.
	
	We also assume that the colloidal particle is spherical, and that it remains in contact with the bubble throughout the growth process. In reality, there may be a partial or complete wetting layer. In Appendix \ref{sec:pressure} we demonstrate that the a thin gap only affects the hydrodynamics at higher order than the global flows. The same is true for the concentration of reaction product in the fluid, since there is no singularity at the contact point and the solution varies continuously when a gap is introduced. At length scales of less than  $\sim\SI{1}{\nano\meter}$, the Rayleigh-Plesset equation \eqref{eq:Lapla} may need to be modified to include effects of adsorption \cite{zhou2022solid}, but these effects would again only be relevant very locally at the contact point and thus a higher-order correction to the findings of this study. A discussion of partial wetting and its consequences for bubble detachment is offered in Appendix \ref{sec:bubbleBgone}.
	
	Finally we assume, also for mathematical simplicity, that the colloid is radially symmetric and all parts of its surface equally participate in the production of solute. As we will show, the gradients of the concentration field surrounding the bubble determine the bubble growth rate, and so the concentration boundary condition on the far side of the colloid can reasonably be expected to be of limited importance. However, since the net flux of solute produced by the colloid will appear as an important parameter, $\mathcal{Q}$, this parameter will need to be rescaled when a particle with lower active surface coverage is considered.
	}
	
 	\subsection{Derivation of the model equations}
 	
 	\begin{figure}[t]
 		\centering
 		\includegraphics[width=0.5\textwidth]{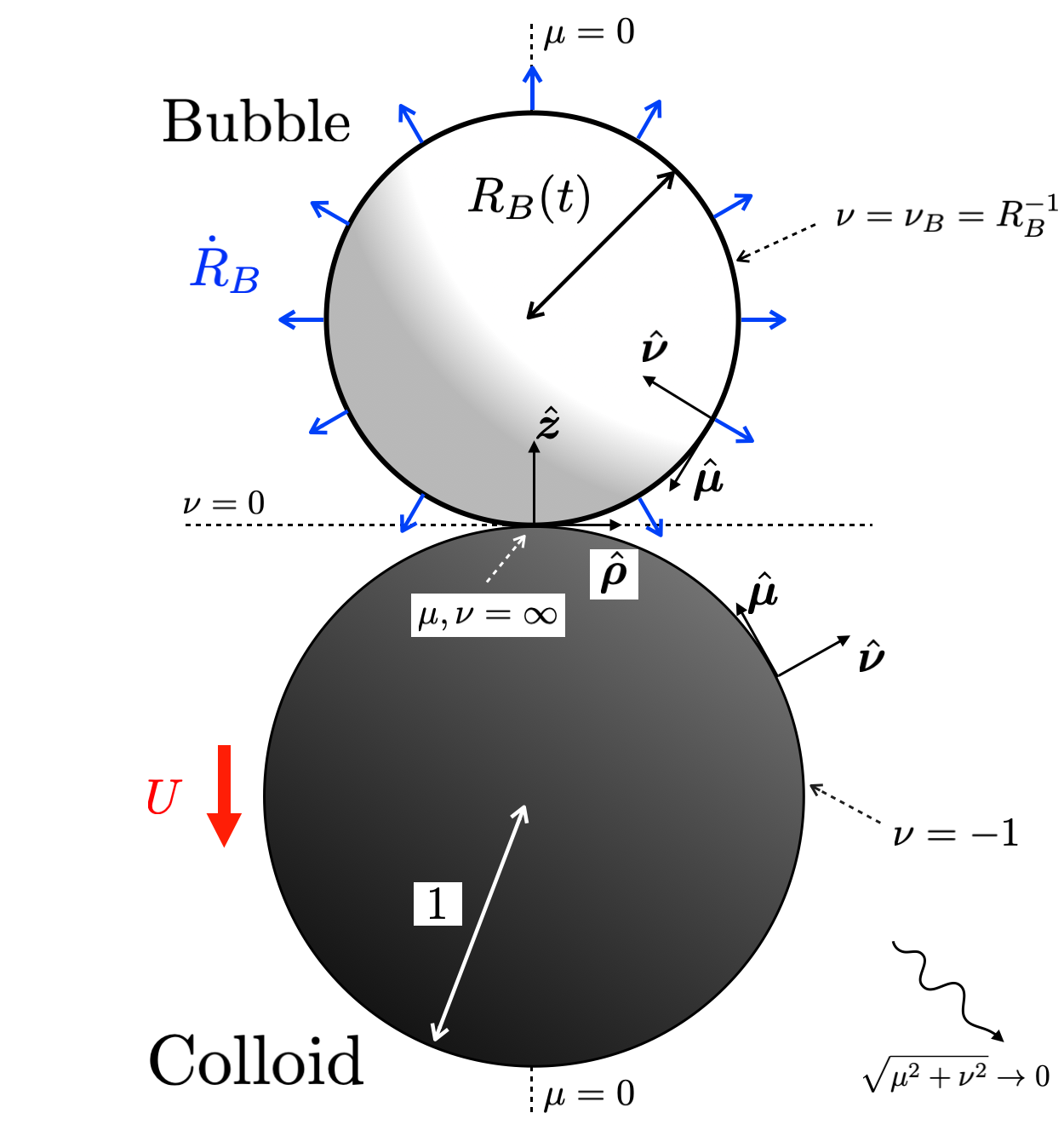}
 		\caption{Growth of a spherical bubble attached to a spherical colloidal particle: sketch of the geometry in coordinates scaled by the colloid radius $R_c$. The bubble (light sphere) has radius $R_B(t)$ and grows at a rate $\dot{R}_B$ while the colloidal particle (dark sphere) has unit radius. The whole system translates with a velocity $U$ along the axis of symmetry ($\mu=0$), defined positive when the bubble pushes the colloid. Both bubble and colloid are assumed spherical and touch at the origin of a cylindrical coordinate system $\{\rho,z\}$, which is equivalent to the point at infinity in tangent-sphere coordinates $\{\mu,\nu\}$.}\label{fig:sketch}
 	\end{figure}

	\revision{We use tangent sphere coordinates $\{\mu,\nu,\theta\}$ as defined in Appendix \ref{sec:coords} to conveniently describe the geometry, with surfaces of constant $\nu$ defining spheres that touch at the origin. The bubble is considered to be the sphere defined by $\nu=\nu_B:=R_B^{-1}$, and the colloidal particle of radius $R_c$ to be the sphere defined by $\nu=-\nu_c:=-R_C^{-1}$. A sketch of the geometrical setup is provided in Fig.~\ref{fig:sketch}. Note that in the figure we scale lengths by the colloid radius $R_c$, in line with the dimensionless formulation of the problem that will be introduced in section \ref{sec:dimless}.}
		
	The colloidal particle acts as a catalyst for a chemical reaction. The  concentration of the reaction product, $c$, satisfies the quasi-steady diffusion equation as
	\begin{align}\label{eq:diffusion}
		D\nabla^2 c=0,
	\end{align}
	where $D$ is the molecular diffusivity. The problem is axisymmetric and the solution $c(\mu,\nu)$ is therefore independent of the azimuthal angle $\theta$ (Fig.~\ref{fig:sketch}). The concentration on the bubble surface is set by Henry's law~\cite{plesset1977bubble}, which states that concentration is proportional to the partial pressure of the corresponding gas inside the bubble, i.e.~$p_\text{bubble}$. Assuming that the bubble contains only the reaction product, Eq.~\eqref{eq:Lapla} leads to
	\begin{align}\label{eq:diffbc1}
		c=K_H^{-1}\left(p^\infty+2\gamma\nu_B\right),\quad\text{at }\nu=\nu_B,
	\end{align}
	where $K_H$ is the constant of Henry's law, which depends on the dissolved gas.

	\revision{As discussed in \ref{sec:assumptions}, we assume that the entire surface participates equally in the reaction.} We model this with \emph{zeroth-order kinetics} (0ok), which assumes a constant catalytic flux $\mathcal{A}$ per unit area that is independent of other factors. This results in a Neumann boundary condition for the concentration product of the form
	\begin{equation}\label{eq:diffbc2_0ok}
		-D\hat{\bm{\nu}}\cdot\bm{\nabla}c = \mathcal{A},\quad\text{at }\nu=-\nu_c. 
	\end{equation}
	A more realistic description is \emph{first-order kinetics} (1ok), which involves a second chemical  solute (the `fuel') of concentration $c$, that is converted into a product (concentration $c$) on the colloid surface at a constant rate. We find that this does not significantly change the $c$ field and only leads to an effective rescaling of global catalytic activity, so a  discussion of this boundary condition is relegated to Appendix \ref{sec:1ok}. 
	
	Finally, the boundary condition far   from the colloidal particle is
	\begin{align}\label{eq:diffbc3}
		c \to c^\infty,\quad\text{as }\sqrt{\mu^2+\nu^2}\to0,
	\end{align}
	where $c^\infty$ is the background concentration of reaction product. Normally, the value of $c^\infty$  is determined by applying Henry's law at the interface with the surrounding atmosphere.
	
 	The instantaneous concentration field for a given bubble radius is determined by the solution to 	Eq.~\eqref{eq:diffusion} along  with the boundary conditions in Eqs.~\eqref{eq:diffbc1}-\eqref{eq:diffbc3}. The instantaneous growth rate of the bubble is then determined by mass conservation,
	\begin{align}\label{eq:diffgrowthode}
		\frac{d}{dt}\left(\frac{4\pi\rho_g}{3}R_B^3\right)=-D\int_{\nu=\nu_B} \hat{\bm{\nu}}\cdot\bm{\nabla}c\, \text{d}S,
	\end{align}
	which states that the rate of change of the bubble mass (left) is equal to diffusive mass flux into the bubble (right). 
	
	Meanwhile, since the Reynolds number is very small, the fluid dynamics surrounding the system are governed by the linear and quasi-steady incompressible Stokes equations~\cite{happel2012low},
	\begin{align}
		\bm{\nabla}p = \eta \nabla^2\bm{u},\quad \bm{\nabla}\cdot\bm{u}=0,
	\end{align}
	where $\eta$ is the dynamic fluid viscosity. The Stokes problem described above is linear with respect to its driving kinematics, the bubble growth velocity $\dot{R}_B$; the resulting instantaneous translation velocity of the bubble-colloid pair, $U(t)$, is linear $\dot{R}_B$, with a constant of proportionality that depends only on the geometry, i.e. the ratio of bubble and colloid radii:
	\begin{align}\label{eq:UtoRB}
		U(t) = f\left(\frac{R_B}{R_c}\right) \dot{R}_B.
	\end{align}
	The function $f$ can be calculated by imposing that the colloid-bubble system remains force-free~\cite{happel2012low}. 
	
	As boundary conditions for the fluid flow, we assume an undisturbed fluid in the bulk (i.e.~$\bm{u} \to \bm{0}$ as $\sqrt{\mu^2+\nu^2}\to0$) and a no slip condition on the colloidal particle. On the bubble, we solve for both cases of a no stress (i.e.~free surface) and a no slip condition, the latter to account for a possible rigidification of the bubble due to surfactants. Combining Eqs.~\eqref{eq:diffgrowthode} and \eqref{eq:UtoRB} with the solution to Eq.~\eqref{eq:diffusion} for the instantaneous concentration field, this determines the 
	 velocity $U(t)$ of the bubble-colloid pair at each point in time.
 	
	\subsection{Dimensionless formulation and sketch of the solution method}\label{sec:dimless}
	
	We now reformulate the problem in dimensionless form by introducing the following scalings for lengths, concentration and time respectively,
	\begin{align}
		\tilde{R}=R_c,\quad \tilde{c}=\frac{2\gamma}{K_HR_c}, \quad \tilde{t}=\frac{K_HR_c^2}{3 \mathcal{R}TD},
	\end{align}
	that is we scale lengths by the colloid radius, concentrations by the typical value on a bubble the same size as the colloid, and time by a natural scale for the diffusive growth described by Eq.~\eqref{eq:diffgrowthode}. As we show in Appendix \ref{sec:app_diff}, the general solution to the diffusion equation, Eq.~\eqref{eq:diffusion}, that decays far from the colloid, Eq.~\eqref{eq:diffbc3}, may be written exactly as
	\begin{align}\label{eq:diffproblem}
		c(\mu,\nu)=\tilde{c}^\infty+(\mu^2+\nu^2)^{1/2}\int_0^\infty \left[A(s)e^{s\nu}+B(s)e^{-s\nu}\right]J_0(s\mu)\, \text{d}s,
	\end{align}
	where $\tilde{c}^\infty=c^\infty K_H R_c/2\gamma$ and the functions $A(s)$ and $B(s)$ are such that the integral converges in the range $\mu\geq 0$, $-1\leq\nu\leq\nu_B$. Here $J_0$ denotes the zeroth order Bessel function of the first kind. For zeroth-order kinetics, the two remaining boundary conditions Eqs.~\eqref{eq:diffbc1} and \eqref{eq:diffbc2_0ok} simplify to
	\begin{align}
		c-\tilde{c}^\infty&=\xi+\nu_B \qquad\text{at }\nu=\nu_B,\label{eq:Diffbc1}\\
		\frac{\partial c}{\partial\nu} &= -\frac{2\mathcal{Q}}{\mu^2+1}\quad\text{at }\nu=-1,\label{eq:Diffbc2_0ok}
	\end{align}
	where we define the dimensionless parameters $\xi$ and $\mathcal{Q}$ as
	\begin{align}
		\xi=\frac{p^\infty R_c}{2\gamma}\left(1-\frac{K_Hc^\infty}{p^\infty}\right),\quad \mathcal{Q}=\frac{\mathcal{A}K_HR_c^2}{2\gamma D}.\label{eq:dimlessparams}
	\end{align}
	Physically, $\xi$ is a dimensionless measure of the saturation of the background fluid, scaled by capillary effects ($=0$ when saturated, $>0$ otherwise). We may therefore interpret $\xi$ as a measure of how `soluble' gas is in the background fluid relative to the bubble.
	Intuitively, a larger value of $\xi$ is therefore conducive to bubble dissolution, and hence limits bubble growth. Meanwhile $\mathcal{Q}$ is interpreted as a scaled catalytic flux out of the colloid, and it is expected that a large value of $\mathcal{Q}$ will be conducive to bubble nucleation and growth.

	Applying the boundary conditions Eqs.~\eqref{eq:Diffbc1}-\eqref{eq:Diffbc2_0ok} to Eq.~\eqref{eq:diffproblem} leads to a regular second-order ODE boundary value problem that can be solved numerically for given values of the parameters $\mathcal{Q}$ and $\xi$. The solution for $A$ may then be substituted into the mass conservation equation, Eq.~\eqref{eq:diffgrowthode}, which simplifies to
	\begin{align}\label{eq:diffgrowthodeND}
		\frac{dR_B}{dt}=-\frac{1}{R_B+\tfrac{3}{2}\zeta R_B^2}\int_0^\infty A(s)\, \text{d}s,
	\end{align}
	where we have introduced the dimensionless group $\zeta=\xi+\tilde{c}^\infty = p^\infty R_c / 2\gamma$. As a corollary we have the condition for a steady state,
	\begin{align}\label{eq:statpoint}
		\int_0^\infty A(s)\, \text{d}s=0,
	\end{align}
	which is satisfied when there is no net flux of gas in or out of the bubble. A detailed derivation of the equations for this diffusive problem and their solution is provided in Appendix \ref{sec:app_diff}.
	
	For the fluid mechanics we use a Stokes streamfunction $\Psi$ with the ansatz
	\begin{align}\label{eq:genstokes_main}
		\Psi=\Psi_s+\frac{\mu}{\left(\mu^2+\nu^2\right)^{3/2}}\int_0^\infty\left[\left(\tilde{A}+\tilde{C}\nu\right)\sinh{s\nu}+\left(\tilde{B}+\tilde{D}\nu\right)\cosh{s\nu}\right]J_1(s\mu)\,\text{d}s,
	\end{align}
	where $\Psi_s\propto \dot{R}_B$ is a singular contribution that accounts for expansion of the bubble, and where the coefficients $\tilde{A}(s)$ to $\tilde{D}(s)$ are found by applying the  hydrodynamic boundary conditions on bubble and colloid (they are thus implicitly a function of the instantaneous bubble radius $R_B$). 
	
	We additionally require that the bubble-colloid system remains force-free at all times, which may be shown to reduce to the condition $\int_0^\infty s \tilde{B}(s)\, \text{d}s=0$ (see Appendix \ref{sec:app_fluid}). Since the boundary conditions are linear in both $U$ and $\dot{R}_B$, the same is true for $\Psi$, and so we may split the full solution for the flow as the linear superposition of a `motility' problem $\propto U$ and a `growth' problem $\propto \dot{R}_B$. The function $f$ in Eq.~\eqref{eq:UtoRB} is then given by
	\begin{align}
		f(R_B) = -\frac{\int_0^\infty s \tilde{B}_\text{gro}(s)\, \text{d}s}{\int_0^\infty s \tilde{B}_\text{mot}(s)\, \text{d}s}.
	\end{align}
	Full details, including entirely analytical expressions for the functions $\tilde{B}_\text{mot}(s)$ and $\tilde{B}_\text{gro}(s)$, are provided in Appendix \ref{sec:app_fluid}. 
	
	\subsection{Parameter estimates}\label{sec:paramestimates}

	In order to provide an \revision{experimentally relevant} model, it is essential to identify the typical ranges and variability of the relevant dimensionless parameters in practical conditions. We take $\gamma\approx\SI{7.2e-2}{\newton\per\metre}$ as for an air-water interface, and $p^\infty \approx \SI{1.0e5}{\pascal}$ for atmospheric pressure. The gas constant is fixed as $\mathcal{R}\approx\SI{8.3}{\joule\per\kelvin\per\mol}$ and at room temperature $T\approx\SI{300}{\kelvin}$ approximately. For oxygen gas dissolved in water at $\SI{25}\degreeCelsius$, $D\approx\SI{2.4e-9}{\metre\squared\per\second}$ and $K_H\approx\SI{7.7e4}{\joule\per\mole}$ are standard values \cite{lide2004crc}. From this, Henry's law calculates a background molarity $c^\infty\approx\SI{2.7e-1}{\mole\per\metre\cubed}$, which is relatively large due to the percentage of oxygen in the atmosphere, which is well known to be $21\% \left(= K_Hc^\infty/p^\infty\right) $ \cite{lide2004crc}. 
	
	For the activity $\mathcal{A}$, it is harder to find an estimate due to strong variability of  conditions and reactions driving the propulsion in the relevant experiments. In the limit of zero Damk\"ohler number (i.e.~when first-order kinetics reduce to zeroth-order), $\mathcal{A}$ is given by $\mathcal{A}\approx kc_f^\infty$, which is the product of reaction rate $k$ and bulk reaction fuel molarity $c_f^\infty$. 
	 In the classical platinum-catalytic $\rm{H_2O_2}$ decomposition we can assume a molarity $c_f^\infty\approx\SI{2.9e2}{\mole\per\metre\cubed}$ (corresponding to 1 wt\% $\rm{H_2O_2}$, a low estimate) and $k\approx\SI{4.1e-5}{\metre\per\second}$ \cite{yu2019porosity,chang2020hydrogen}. Finally,  we consider colloid radii $R_c$ between $\SI{1}{\micro\meter}$ and $\SI{50}{\micro\meter}$, as relevant to all published experiments to do (the colloidal size is the experimentally most variable parameter). 

	Combining all of these, we find the typical range for $\xi$ is $\mathcal{O}(1)-\mathcal{O}(10)$, while $\mathcal{Q}$ is in the range $\mathcal{O}(10)-\mathcal{O}(10^3)$. Furthermore, the growth time scales as $\tilde{t}\approx\mathcal{O}(10)-\mathcal{O}(10^3)\,\si{\milli\second}$, which is consistent with experimental observations~\cite{manjare2012bubble,yang2017peculiar}. Since $\xi$ scales as $\sim R_c$, and $\mathcal{Q}$ and $\tilde{t}$ scale as $\sim R_c^2$, larger values apply to larger colloids, and due to the quadratic dependence on $R_c$, the value of $\mathcal{Q}$ depends quite sensitively on the colloid size. As we shall see in  \ref{sec:diffsol}, this is the fundamental reason why bubble propulsion is limited to sufficiently large colloids.
	
	In the development of our model we assumed that the bubble grows in a quasi-steady fashion, i.e.~we neglected advective terms in Eq.~\eqref{eq:diffusion}. This is appropriate if the P\'eclet number Pe is small, i.e.~if the time scale of diffusive transport is much shorter than the time scale of fluid flow. As we see from Eq.~\eqref{eq:UtoRB}, the time scale of fluid flow is linked to the time scale of bubble growth, which allows us to justify this assumption \emph{a posteriori} by demonstrating that it is self-consistent with its prediction. Quantitatively this is the requirement that
	\begin{align}
		\frac{R_c^2}{D}\ll \frac{K_HR_c^2}{3 \mathcal{R}TD} \quad\Leftrightarrow\quad \text{Pe}\equiv\frac{3 \mathcal{R}T}{K_H}\ll 1
	\end{align}
	where we have assumed diffusion on the scale of the colloid. 
	With our parameter estimates, we obtain 	 $\text{Pe}\approx 0.097$, 
	  which is indeed small compared to unity, and even more so when diffusion on length scales smaller than the colloid is concerned. Thus the quasi-steady assumption is approximately valid and advective transport only leads to a small correction. It is important to note however that for gases other than oxygen, $K_H$ can be substantially different. While for other common gases such as $\rm{N_2}$ and $\rm{H_2}$ we also have $\text{Pe}\ll 1$, this is not the case e.g.\ for $\rm{CO_2}$ ($\text{Pe}\approx2.4)$ or $\rm{NH_3}$ ($\text{Pe}\approx750)$~\cite{lohse2015surface}, in which case it would be essential to solve the fully coupled nonlinear model for the transport of the reaction product.


	\section{Analysis of bubble growth and propulsion dynamics}\label{sec:results}
 
	\subsection{Solute concentration fields and stationary points of the growth dynamics}\label{sec:diffsol}
	
	\begin{figure}[t]
		\centering
		\begin{subfigure}{0.35\textwidth}
			\includegraphics[width=\textwidth]{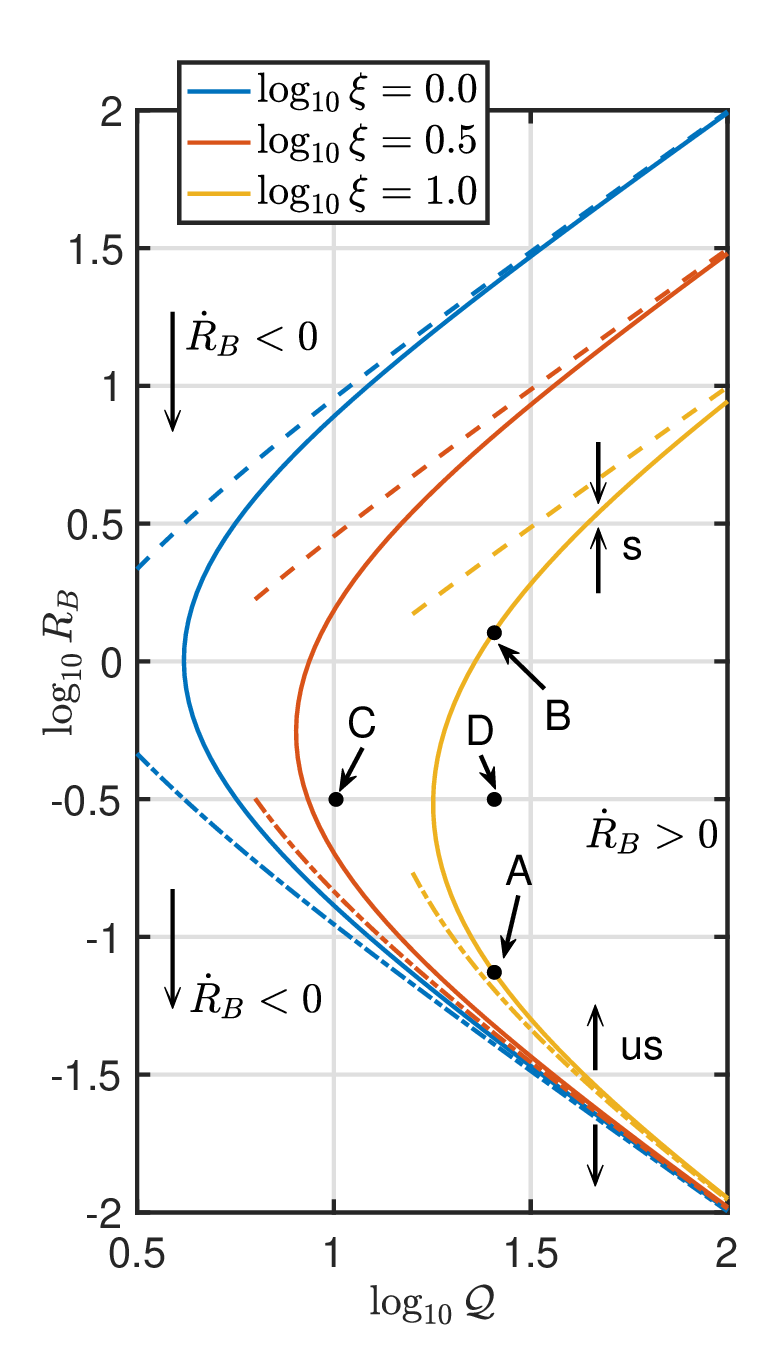}
		\end{subfigure}
		\hfill
		\begin{subfigure}{0.64\textwidth}
			\includegraphics[width=\textwidth]{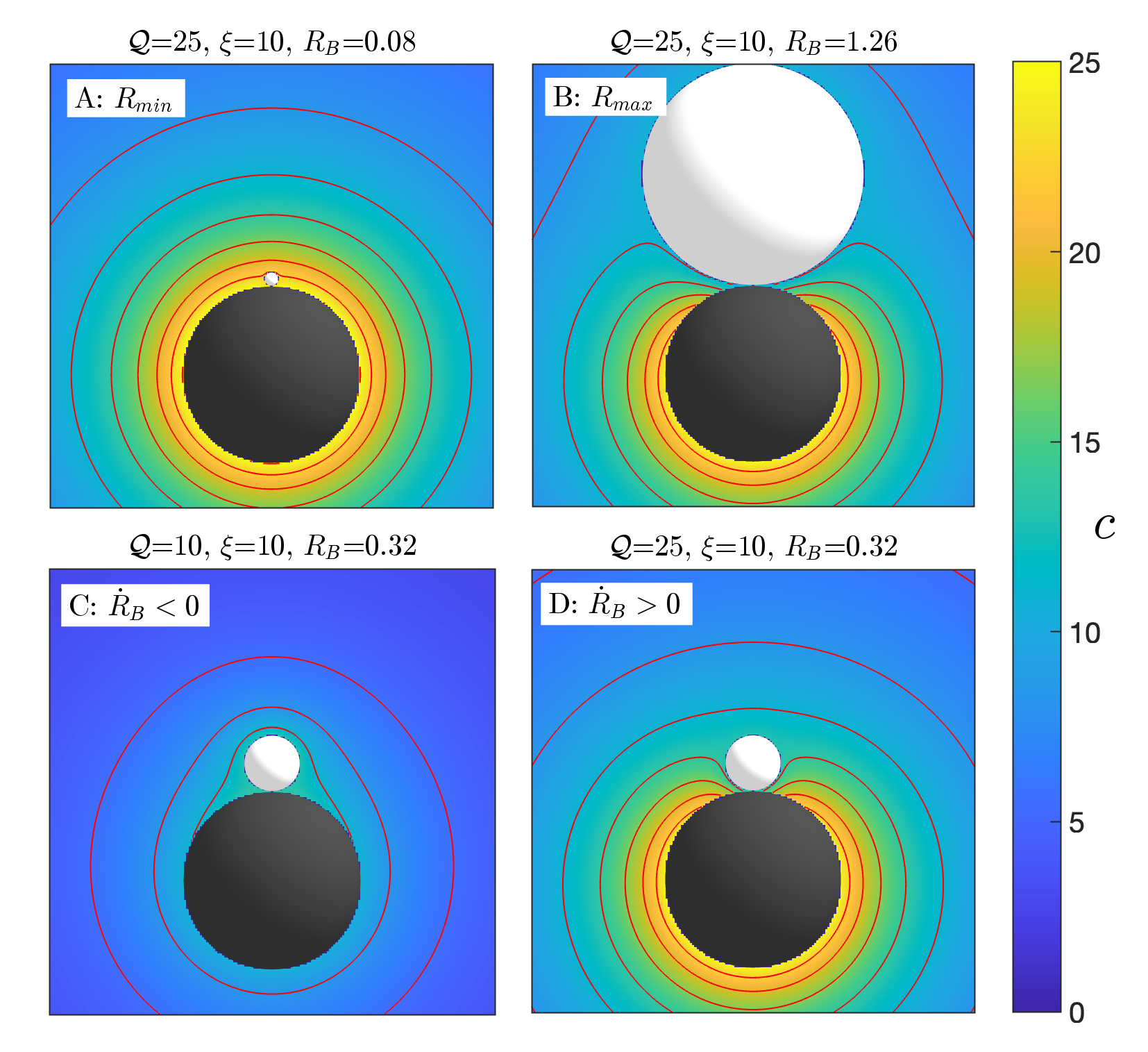}
		\end{subfigure}
		\caption{Left: stationary points of the bubble growth dynamics (solid lines) as a function of the parameters $\mathcal{Q}$ and $\xi$. Dashed and dash-dotted lines indicate theoretical estimates for $R_\text{max}$ (stable) and $R_\text{min}$ (unstable) respectively. Right: concentration fields with $c$-contours in red, the colloid in black, and the bubble in white, for 4 points in $R_B$-$\mathcal{Q}$ space with $\xi=10$.}\label{fig:SteadyDiff}
	\end{figure}
	
	We first solve Eq.~\eqref{eq:diffusion} for the instantaneous concentration field as a function of the dimensionless parameters $\mathcal{Q}$, $\xi$ and the bubble radius $R_B$, and identify the stationary points of the bubble size by means of the condition in Eq.~\eqref{eq:statpoint}. The results are shown in Fig.~\ref{fig:SteadyDiff}. For small values of $\mathcal{Q}$, the bubble always shrinks to zero. Above a critical activity, a saddle-node bifurcation leads to the existence of three regimes with two fixed points such that
	\begin{align}
		\text{bubble shrinks} < R_\text{min} < \text{bubble grows} < R_\text{max} < \text{bubble shrinks}.
	\end{align}
	At the critical point we have approximately that $\mathcal{Q}_\text{crit.}\sim\xi^{0.3}$ and  $R_B\sim\xi^{-0.5}$. 
	
	The behaviour of this system can be understood from a theoretical point of view by considering the competing effects that lead to bubble growth and shrinkage. Small bubbles have high capillary pressure, which enforces a high concentration on their surface through Henry's law. Only if an even larger concentration around the bubble can be sustained through catalytic flux from the colloid can there be bubble growth. Hence, the lower fixed point is achieved when there is a balance between these effects, i.e.
	\begin{align}\label{eq:minpred}
		\mathcal{Q} \sim R_\text{min}^{-1} + \xi \quad\Rightarrow\quad R_\text{min}\sim \left(\mathcal{Q}-\xi\right)^{-1}.
	\end{align}
	For more details of this calculation, we refer to Appendix \ref{sec:exsols}. For a small bubble to grow and propel the colloid, $R_\text{min}$ needs to be sufficiently small. Since $\mathcal{Q}\sim R_c^2$ and $\xi\sim R_c$, this explains quantitatively why bubble propulsion is only observed for strongly catalytic colloids~\cite{manjare2012bubble}. 
	
	In the opposite case, the bubble reaches its final size when influx from the colloid balances with outflux due to diffusive dissolution. This corresponds to the condition
	\begin{align}\label{eq:maxpred}
		\mathcal{Q} \sim 1+\xi R_\text{max}  \quad\Rightarrow\quad R_\text{max}\sim\frac{\mathcal{Q}-1}{\xi},
	\end{align}
	where the origin of these scalings is also discussed in Appendix \ref{sec:exsols}.
	
	In Fig.~\ref{fig:SteadyDiff} we plot the numerically determined stationary bubble sizes (solid lines) against the theoretical predictions from Eqs.~\eqref{eq:minpred}-\eqref{eq:maxpred} (dashed lines) as a function of the dimensionless groups $\mathcal{Q}$ and $\xi$, and find that there is excellent agreement when $\mathcal{Q}\gg\mathcal{Q}_\text{crit.}(\xi)$. The disagreement becomes significant near the bifurcation point, when bubble and colloid are of roughly the same size and geometric details become important.
	
	\subsection{Relationship between bubble growth rate and propulsion} \label{sec:fluids}
	
	\begin{figure}[t!]
		\centering
		\includegraphics[width=0.5\textwidth]{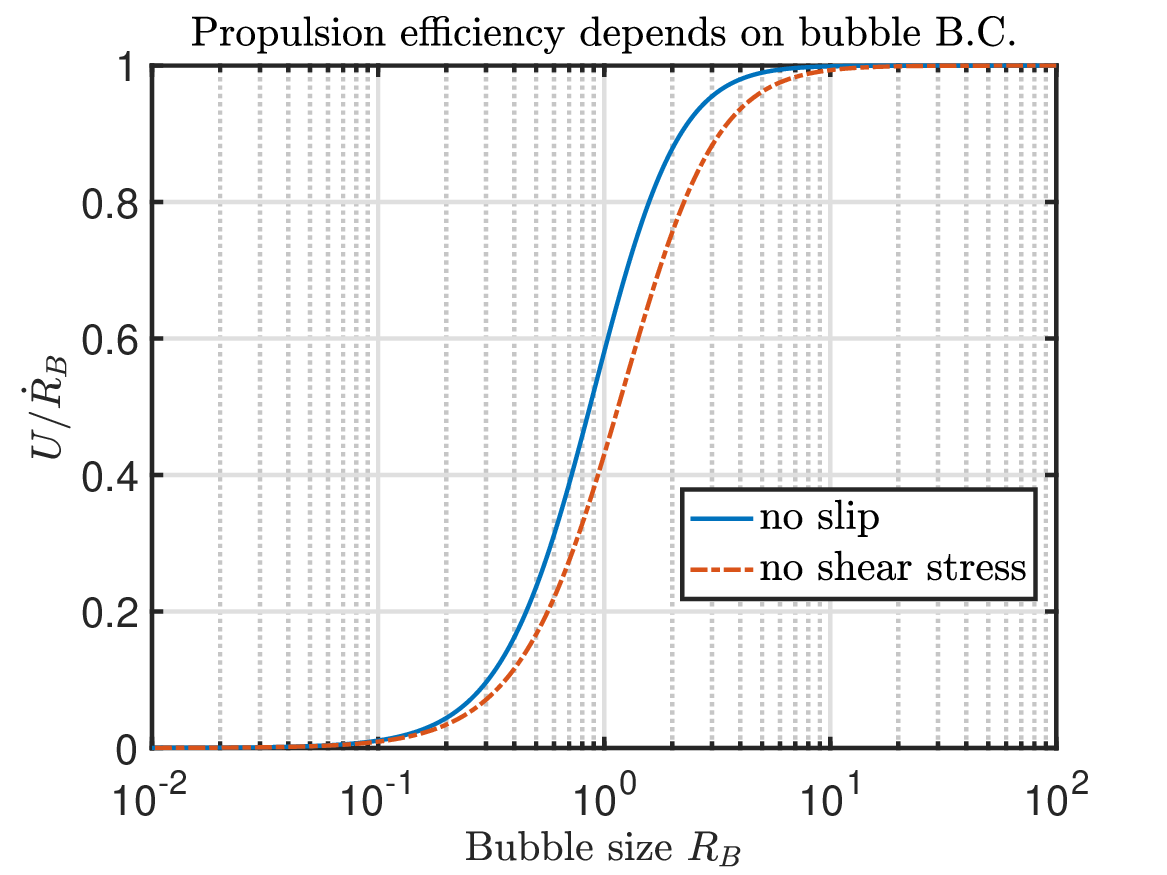}
		\caption{Ratio of propulsion speed, $U$, to bubble growth rate, $\dot R_B$, for no slip (blue solid line) and no shear boundary conditions (red dash-dotted line) as a function of bubble size.}\label{fig:uRB}
	\end{figure}

	Focusing next on propulsion, we now calculate the function $f$ in Eq.~\eqref{eq:UtoRB} for a no slip condition on the colloid, a flow decaying far away on the system, and two different boundary conditions on the bubble: a `rigid' bubble with no slip as on the colloid, and a `shear-free' bubble that supports no tangential shear stress. While the second condition is canonical for an air-water interface in most contexts, it is a well-known fact that surfactants and impurities can lead to a rigidification of these interfaces under experimental conditions \cite{lohse2015surface,rodriguez2015generation}. Since it is not entirely clear which boundary condition applies for this system, and they differ slightly in their physical predictions, we include both in our discussion below.

	The results are illustrated in Fig.~\ref{fig:uRB}, showing $f = U/\dot{R}_B$ as a function of $R_B$ for no slip (blue solid line) and no shear stress (red dash-dotted line) boundary conditions. In both cases, the ratio of propulsion speed to bubble growth rate $U/\dot{R}_B$ increases monotonically with the bubble size, so for a fixed growth rate, large bubbles are more efficient than small ones at pushing the colloid. While the mathematical expressions for the drag and propulsion coefficients are very complicated, the asymptotic behaviour for large and small bubbles can be understood intuitively. As $R_B\to\infty$ ($\nu_B\to0$), both the drag and propulsion scale as $R_B$ and we recover the drag coefficients for a single sphere in Stokes flow, i.e.~$6\pi\eta R_B$ or $4\pi\eta R_B$ depending on the boundary condition (no slip or no shear). In this limit, the colloid's contribution to drag is negligible and it is just pushed outwards with velocity $U=\dot{R}_B$ by the growing bubble that remains stationary itself. Conversely, as $R_B\to 0$ the drag asymptotes to that of an isolated colloid, $6\pi\eta$, while the propulsion force decays as $R_B^{-2}$. This is line with expectations of the force exerted by a point source on a rigid sphere, which scales as $Q/d$ with $Q$ the source strength and $d$ the distance to the colloid centre~\cite{chamolly2020stokes}. 
	Interestingly, it makes no difference in this limit whether the boundary condition on the bubble is no slip or shear-free. The difference is most pronounced when bubble and colloid are of equal size, and because a rigid surface has more drag the propulsion is up to $50\%$ more efficient when the action of surfactants is strong.

	\subsection{Time-dependent propulsion dynamics}\label{sec:combined}
	
	\subsubsection{Evolution of a single growth cycle and comparison with experiments}\label{sec:comp_exp}
		
	\begin{figure}[t!]
		\centering
			\includegraphics[width=0.5\textwidth]{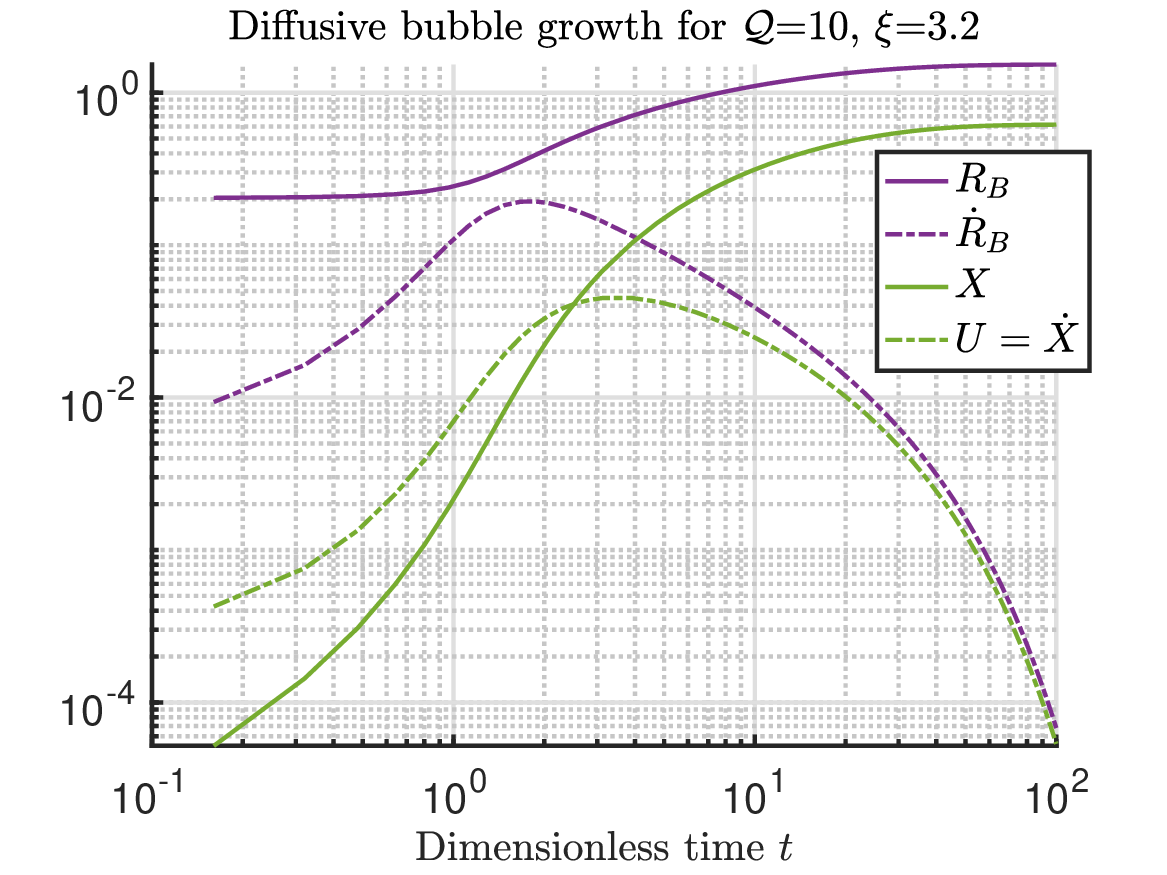}
		\caption{Example of propulsion over one bubble growth cycle for intermediate values of the parameters, $\mathcal{Q}=10$ and $\xi=10^0.5$. Shown are the bubble radius $R_B$ (purple) and travelled distance $X$ (green), as well as their time-derivatives (dashed lines) against logarithmic time. Growth and propulsion rates increase rapidly on very short time scales, peak at intermediate ones, and decay as the bubble asymptotically reaches its stable equilibrium size at long times.}\label{fig:pdEx}
	\end{figure}
	
	We are now in a position to combine all of our results to find the colloid propulsion speed, $U$, and distance travelled, $X$, over one bubble growth cycle as a function of the dimensionless parameters $Q$ and $\xi$, which we recall represent catalytic flux magnitude and solubility of gas in the background fluid relative to the bubble. 
	
	To learn about the generic behaviour of the system, we initialise the bubble at a size slightly above the unstable stationary point $R_\text{min}$, calculate the growth rate according to Eq.~\eqref{eq:diffgrowthodeND} and the rate of displacement $\dot{X}=U$ from Eq.~\eqref{eq:UtoRB}, and integrate the system forward in time using \revision{first-order explicit} Euler time-stepping. A typical example of the resulting dynamics is shown in Fig.~\ref{fig:pdEx}, where we display the time-variation of ${R}_B$, $\dot{R}_B$, $X$ and $\dot X$.  We note that it follows from time-integration of Eq.~\eqref{eq:UtoRB}, that the displacement $X$ is slaved to the instantaneous bubble size. Since small bubbles propel the colloid insignificantly, the bubble nucleation size has little influence over the final displacement. Hence, the final value of $X$ is determined, to within a good approximation, only by the final bubble size and the hydrodynamic boundary condition on the bubble. Consequentially, the only variation in the dynamics shown in Fig.~\ref{fig:pdEx} for different values of the parameters are in the variation of the final bubble size (discussed previously in section \ref{sec:diffsol}), and the time scales of growth.
	
	The logarithmic representation of the growth rate reveals three distinct regimes at short, intermediate and long times. At very short times, when the bubble size is close to $R_{\rm min}$, velocities and displacements are \ small. However, since the deviation from an unstable fixed point generically follows exponential growth~\cite{strogatz2018nonlinear}, the dynamics transition rapidly to an intermediate regime in which the growth rate $R_B$ peaks. In this regime the bubble typically crosses from being  smaller than the colloid to being larger, which also maximises the colloid velocity $U$. At long times, the approach to the stable fixed point $R_\text{max}$ is again exponential, and no significant propulsion occurs.
	
	The striking prediction of our diffusive growth model is thus that parameter values that allow for the growth of very small bubbles simultaneously allow these bubbles to grow orders of magnitude larger than the colloid, unless the experiment is performed in a narrow region of the parameter space in which $\xi$ is large and $\mathcal{Q}$ not much larger than its critical value. This was illustrated in Fig.~\ref{fig:SteadyDiff} and rationalised with scaling arguments for the stationary points in Eqs.~\eqref{eq:minpred}-\eqref{eq:maxpred}. Due to the scalings of these dimensionless parameters with the colloid radius, Eq.~\eqref{eq:dimlessparams}, this is difficult to achieve in practice unless the experiment is carried out in a high-pressure environment. Indeed, both bubbles that are much larger than the colloid and a slow asymptotic approach to a stable equilibrium are typically not observed in experiments, even though the catalytic activity is clearly strong enough to allow for the nucleation of small bubbles~\cite{manjare2012bubble}.
	
	Aiming to address this apparent paradox, we considered both more realistic models for the catalytic reaction, and more detailed descriptions of the contact physics between bubble and colloid. Neither have brought to light any credible intrinsic mechanism that leads to these observations. Accounting for the depletion of reaction fuel with a non-zero Damköhler number leads to an effective decrease of the activity parameter $\mathcal{Q}$, but not to a different scaling law for the maximal bubble radius (see section \ref{sec:1ok}). Likewise, an analytical asymptotic solution for the pressure at the contact reveals that there is no divergent hydrodynamic force pulling the system apart, and both a scaling analysis and a simplified model with a non-singular contact demonstrate that bubble is too small for entrainment or bouyancy to be significant (see sections \ref{sec:pressure} and \ref{sec:bubbleBgone}). We  therefore conjecture that the premature disappearance of bubbles that was reported in Ref.~\cite{manjare2012bubble} is in fact due to an external perturbation, most likely the coalescence of the bubble with the air-water interface of the experimental setup. This idea  is in fact supported by analogous observations in later work~\cite{yang2017peculiar}. 
	 
	While a detailed model of bubble growth in confinement is beyond the scope of this study,  in the context of micro-rockets it has been shown numerically and experimentally that confining bubble growth is beneficial in achieving efficient propulsion~\cite{gallino2018physics,gao2012hydrogen}. 
	Interpreting the bubble disappearance at the air-water interface as a confinement effect, we note that this is another example of this principle, albeit indirectly by limiting `natural' bubble growth to the point in which it is efficient in propelling the colloid forwards.

	In order to effectively make a prediction for experimental systems we thus need introduce a phenomenological cut-off for the bubble size into our model. Since the colloid sets a natural scale in terms of confinement, we choose $R_\text{max}=1$ for this; note that it could also be larger if the colloid is in a deep layer~\cite{manjare2012bubble}.

	\subsubsection{Dependence of time-averaged propulsion on catalytic activity and background saturation}
	
	\begin{figure}[t]
	\centering
	\begin{subfigure}{0.49\textwidth}
		\includegraphics[width=\textwidth]{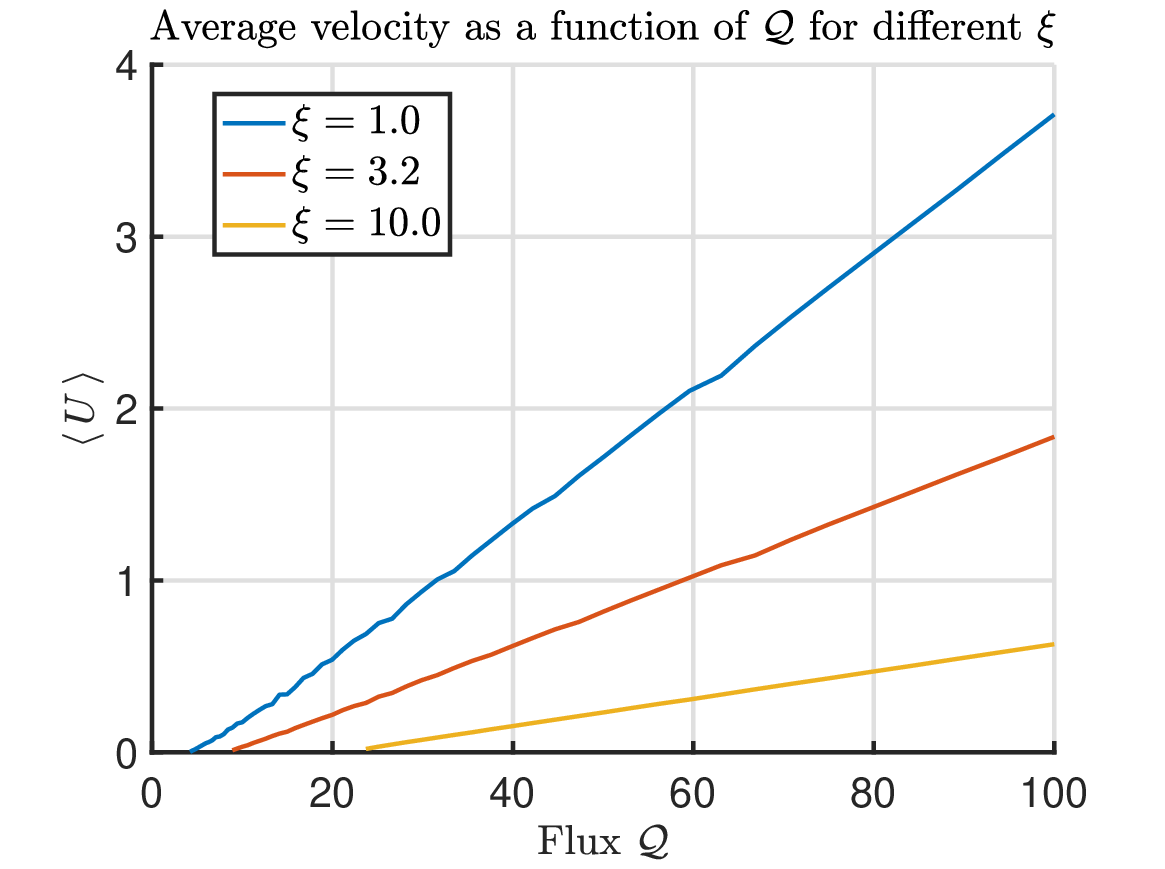}
	\end{subfigure}
	\hfill
	\begin{subfigure}{0.49\textwidth}
		\includegraphics[width=\textwidth]{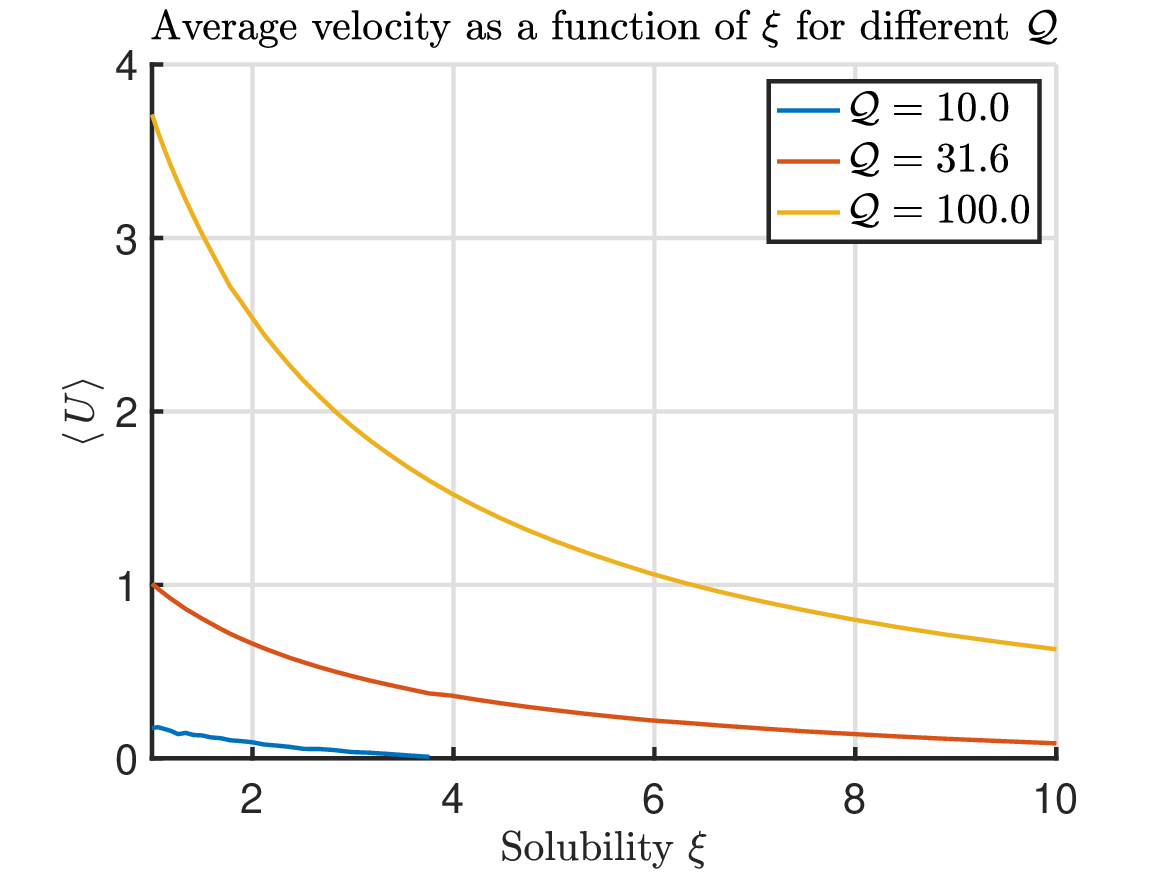}
	\end{subfigure}
	\caption{Simulated average propulsion speed $\langle U \rangle$ over one bubble growth cycle for rigid boundary conditions on the bubble surface, as a function of flux $\mathcal{Q}$ (left) and background solubility $\xi$. Above a certain $\xi$-dependent threshold,  $\langle U \rangle$ increases linearly with $\mathcal{Q}$. Conversely, the propulsion speed decreases as the solubility in the bulk increases.}\label{fig:U}
	\end{figure}
	
	From a practical point of view, an interesting question concerns the propulsive efficiency of the system as a function of its parameters. As established in section \ref{sec:fluids}, the total distance travelled, $X$, depends only on the final bubble size. Fixing this at the cut-off $R_B=1$ leads to a prediction of the colloid displacement over one growth cycle as
	\begin{align}
		X_\text{rigid bubble}/R_C \approx 0.25 ,\qquad X_\text{shear-free bubble}/R_C \approx 0.18,
	\end{align}
	independently of the parameters $\mathcal{Q}$ and $\xi$.
	 
	 Another quantity of interest is the time scale on which this displacement is achieved, or equivalently, the average speed of the colloid. We define this average $\langle U \rangle$ by the distance $X$ travelled divided by the time required to reach the cut off size from a small perturbation away from the unstable fixed point $R_\text{min}$. The dependence of this average velocity on the two relevant parameters ($\mathcal{Q}$ and $\xi$) is displayed in Fig.~\ref{fig:U}. Due to the rapidness of the initial exponential growth phase, these results are robust towards the exact size of the initial perturbation. For reference, we recall from our parameter estimates in section \ref{sec:paramestimates} that $U=1$ in dimensionless units corresponds to the physical velocity
	\begin{align}
		U  =  \frac{\SI{2.3e-4}{\metre\per\second}}{R_c\, \si{\per\micro\metre}},
	\end{align}
	which is on the order of $\SI{1}{\milli\meter\per\second}$. The dependence on the catalytic activity of the colloid $\mathcal{Q}$ is approximately linear above a cut-off $\mathcal{Q}_\text{crit.}$, below which the bubble does not grow (Fig.~\ref{fig:U}, left). Close to this cut-off there is a small deviation from linear behaviour as the slow asymptotic approach to $R_\text{max}$ becomes significant. Unsurprisingly, a larger activity leads to faster propulsion. Within the scope of our quasi-static model this growth is unbounded -- in practice fast velocities will eventually lead to non-negligible P\'eclet and Reynolds number that regularise this behaviour. Additionally, a stronger solubility of gas in the fluid $\xi$ has a negative effect on propulsion (Fig.~\ref{fig:U}, right). 
	
	\subsection{Conservation of momentum during inertial bubble collapse}\label{sec:inertia}
	
	\begin{figure}
		\centering
		\includegraphics[width=0.8\textwidth]{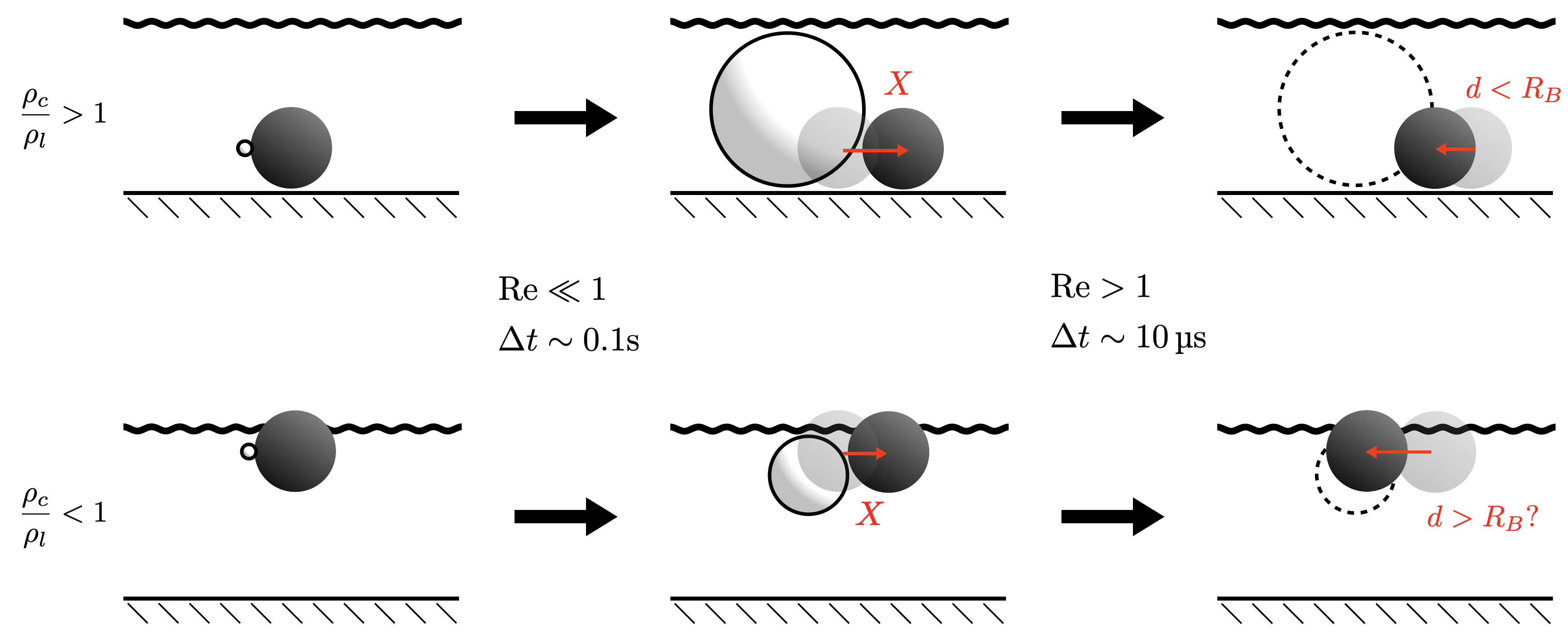}
		\caption{Thought experiment on bubble retraction due to inertial collapse. Based on momentum considerations, the retraction distance $d$ is shorter than the propulsion distance if the colloid is heavy. But if the colloid is lighter than water, a net pulling by the bubble \revision{might} be achieved if inertial retraction outweighs pushing due to bubble growth.}\label{fig:retraction}
	\end{figure}
	
	Finally, we briefly address the second phase of the propulsion cycle, which is bubble collapse. As established in section \ref{sec:comp_exp}, the bubble size is limited in practice by the depth of the fluid layer in which the colloid is immersed, and \revision{so it is likely} that the coalescence with the top interface leads to its disappearance. This process is very fast, and so for a brief moment the dynamics become inertial, and the colloid retracts a short distance~\cite{manjare2012bubble}. While a numerical simulation of this process \revision{would require faithful modelling of the geometry and non-linear flows, and} is beyond the scope of this paper, we would like to present a simple thought experiment based on conservation of linear momentum that suggests interesting dynamics might happen for a colloid that is lighter than water.
	
	Specifically, if we call the retraction distance $d$, we can consider conservation of linear momentum before and after the bubble collapse to find $d$ as a function of the bubble radius before collapse, and of the densities of the colloid and the surrounding fluid. Since there is no net momentum change, the centre of mass of the whole system remains approximately stationary. The colloid is displaced a distance $d$ backwards, while fluid is displaced a distance $2R_B-d$ forwards \revision{(assuming no inflow from the sides)}. Denoting by $\rho_c$ and $\rho_l$ the densities of the colloid and the surrounding fluid respectively we then have that
	\begin{align}
		d\rho_c -(2R_B-d)\rho_l =0 \quad\Rightarrow\quad d=\frac{2R_B}{1+\tfrac{\rho_c}{\rho_l}}.
	\end{align}
	This is illustrated by a sketch in Fig.~\ref{fig:retraction} (top row). Of course, this simplified argument ignores many important factors such as directionality of flows, time-dependence and confinement effects, and should therefore only be understood as a rough estimate for $d$ as a function of the density ratio $\rho_c/\rho_l$. Within these limitations, we see then that for a heavy colloid (i.e.~one with \revision{$\rho_c\gg \rho_l$}) we have $d\ll R_B$, and so the retraction distance is much smaller than the final bubble radius, which is itself on the same order as the propulsion distance. As a result, the net displacement is such that the bubble pushes the colloid over one growth cycle.
  
	Conversely, one could imagine a scenario where $\rho_c<\rho_l$, where the simple argument predicts that $d>R_B$. This would correspond to a floating colloid, and it is illustrated in Fig.~\ref{fig:retraction} (bottom row). This suggests that a floating colloid may actually be net \emph{pulled} by the bubble, since the inertial retraction due to collapse could outweigh non-inertial propulsion due to growth.Whether this is possible in practice remains to be seen, but we would be excited to see this hypothesis tested experimentally.
		
\section{Discussion}\label{sec:discussion}
 
In this work we have developed a combined diffusive-hydrodynamic model for the propulsion of a catalytic colloid by a growing gas bubble. We have identified key non-dimensional parameters of the system and their relation to experimentally accessible quantities, in particular the colloid size. With this, we have gained new understanding of the conditions which allows the nucleation of small bubbles that mark the departure from purely phoretic modes of propulsion into a regime where bubble growth allows colloids to reach much higher speeds. In particular, we identified that conditions that allow small bubbles to grow simultaneously allows them to grow much larger than the colloid, unless an external event or forcing leads  to their detachment or disappearance. Finally, we combine all of our results to find a prediction for the propulsion speed of the system as a function of the catalytic activity of the colloid, and the solubility of gas in the surrounding medium. We conclude by  predictions for an as yet unrealised phenotype in which a light catalytic colloid can surf on water being pulled by the collapse of bubbles that it generates itself.

\revision{As discussed in section \ref{sec:assumptions}, our results are obtained under a number of simplifying assumptions in an effort to analyse the fundamental mechanisms on the simplest possible system. We purposefully do not address the question of bubble nucleation, nor do we consider consider multiple bubbles due to the mathematical complexity of the fluid mechanics. Instead, we demonstrate how fundamental physical ingredients can be combined to give novel predictions to be tested quantitatively in experiments. More advanced modelling of the dynamics e.g.\ at the contact point will be of interest once the leading-order predictions obtained here have been validated. In particular, nanoporous colloids with potentially sophisticated bubble nucleation and contact line dynamics present an interesting future research direction.}

Our results pave the way for further experimental and theoretical investigations of bubble propulsion. Since the focus has been on gaining an analytical understanding of the fundamental physics, our study has been mostly limited to the geometry of a bubble-colloid pair in isolation. Yet as discussed confinement plays a crucial role by clearing the bubble from the system. Further investigation may reveal the extent to which this process can be optimised, and how the substrate on which heavy colloids are typically located affects propulsive efficiency. 

\begin{acknowledgments}
We would like to thank Javier Rodriguez-Rodriguez for useful discussions. This work was supported by the European Research Council (ERC) under the European Union's
Horizon 2020 research and innovation program (grant agreement no. 714027 to S.M.). For the purpose of Open Access, a CC-BY public copyright licence has been applied by the authors to the present document and will be applied to all subsequent versions up to the Author Accepted Manuscript arising from this submission.
\end{acknowledgments}

\appendix
	\section{Properties of tangent sphere coordinates}\label{sec:coords}
	
	
	We define tangent sphere coordinates $\{\mu,\nu,\theta\}$ in terms of usual cylindrical coordinates $\{\rho,\theta,z\}$ as
	\begin{align}
		\mu=\frac{2\rho}{\rho^2+z^2}, \quad \nu=\frac{2z}{\rho^2+z^2},\quad \theta=\theta.
	\end{align}
	Their ranges are $\mu\geq0$, $\nu\in\mathbb{R}$, and $\theta\in[0,2\pi)$. They are essentially an `inversion' of cylindrical coordinates. As such, $\mu$ and $\nu$ have units of inverse length, and the coordinate origin and point at infinity have been exchanged. Surfaces of constant $\nu$ define spheres with radius $1/|\nu|$ and their centre located on the $z$-axis at $z=1/\nu$. Thus all surfaces of constant $\nu$ touch the plane $z=0$ at the spatial origin, and the sign of $\nu$ determines whether they are located in the half space $z>0$ (for $\nu$ positive), or $z<0$. The surface $\nu=0$ corresponds to the plane $z=0$. Similarly, surfaces of constant $\mu$ define toroids around the $z$-axis with circular cross-sectional radius $1/\mu$. The surface defined by $\mu=0$ is degenerate and corresponds to the $z$-axis. Finally, $\theta$ is the azimuthal angle as in cylindrical coordinates.
	
	The scale factors are
	\begin{align}
		h_\mu=h_\nu=\frac{2}{\mu^2+\nu^2},\quad h_\theta=\frac{2\mu}{\mu^2+\nu^2},
	\end{align}
	and the basis vectors are
	\begin{align}\label{eq:coordconv}
		\hat{\bm{\mu}}&=-\frac{\mu^2-\nu^2}{\mu^2+\nu^2} \hat{\bm{\rho}}-\frac{2 \mu\nu}{\mu^2+\nu^2}\hat{\bm{z}}, \\
		\hat{\bm{\nu}}&=-\frac{2 \mu\nu}{\mu^2+\nu^2}\hat{\bm{\rho}} +\frac{\mu^2-\nu^2}{\mu^2+\nu^2} \hat{\bm{z}}.
	\end{align}
	Hence $\bm{\nabla} =\frac{\mu^2+\nu^2}{2}\left(\hat{\bm{\mu}}\frac{\partial}{\partial\mu}+\hat{\bm{\nu}}\frac{\partial}{\partial\nu}+\hat{\bm{\theta}}\mu^{-1}\frac{\partial}{\partial\theta}\right)$ and in particular
	\begin{align}\label{eq:nugrad}
		\hat{\bm{\nu}}\cdot\bm{\nabla}=\frac{1}{h_\nu}\frac{\partial}{\partial\nu}=\frac{\mu^2+\nu^2}{2}\frac{\partial}{\partial\nu},
	\end{align}
	regardless of the sign of $\nu$.
	
	\section{Detailed solution of the steady diffusive growth model}\label{sec:app_diff}
		\subsection{Zeroth-order kinetics}  
		We assume non-dimensional scalings. Recalling from the main text, we aim to solve the problem
		\begin{align}\label{eq:diffusionapp}
			\nabla^2 c = 0,
		\end{align}
		subject to the boundary conditions
		\begin{align}
			c-\tilde{c}^\infty&=\xi+\nu_B \qquad\text{at }\nu=\nu_B,\label{eq:dbc1app}\\
			\frac{\partial c}{\partial\nu} &= -\frac{2\mathcal{Q}}{\mu^2+1}\quad\text{at }\nu=-1,\label{eq:dbc2app}\\
			c &\to \tilde{c}^\infty\quad\qquad\text{as }\sqrt{\mu^2+\nu^2}\to 0 .\label{eq:dbc3app}
		\end{align}
		Definding $G(\mu,\nu)$ as
		\begin{align}
			c=\tilde{c}^\infty+\sqrt{\mu^2+\nu^2}G(\mu,\nu),
		\end{align}
		the diffusion equation Eq.~\eqref{eq:diffusionapp} reduces to
		\begin{align}
			\frac{1}{4}(\mu^2+\nu^2)^{5/2}\left(G_{\mu\mu}+\mu^{-1}G_\mu+G_{\nu\nu}\right)=0.
		\end{align}
		Thus a solution that satisfies the boundary condition at spatial infinity, Eq.~\eqref{eq:dbc3app}, is determined by a solution for $G$ that remains bounded as $\sqrt{\mu^2+\nu^2}\to 0$. Using separation of variables this yields $G\sim\exp(\pm s\nu)J_0(s\mu)$, where $J_0$ denotes the zeroth order Bessel function of the first kind (a second family of solutions, proportional to the divergent $Y_0(s\mu)$, is discarded). Hence
		\begin{align}
			c(\mu,\nu)=\tilde{c}^\infty+(\mu^2+\nu^2)^{1/2}\int_0^\infty \left[A(s)e^{s\nu}+B(s)e^{-s\nu}\right]J_0(s\mu)\, \text{d}s,
		\end{align}
		for some functions $A(s)$ and $B(s)$ that are such that the integral converges in the range $\mu\geq 0$, $-1\leq\nu\leq\nu_B$. 

		To solve the steady diffusive problem, we need to find the functions $A(s)$ and $B(s)$. The boundary condition on the bubble, Eq.~\eqref{eq:dbc1app}, gives
		\begin{align}
			\frac{\nu_B+\xi}{\sqrt{\mu^2+\nu_B^2}}=\int_0^\infty \left[A(s)e^{s\nu_B}+B(s)e^{-s\nu_B}\right] J_0(s\mu )\, \text{d}s,
		\end{align}
		while the one on the colloid, Eq.~\eqref{eq:dbc2app}, yields
		\begin{align}\label{eq:BC2}
			\frac{-2\mathcal{Q}}{\sqrt{\mu^2+1}}&=\int_0^\infty \left[A(s)\left[s(\mu^2+1)-1\right]e^{-s}+B(s)\left[-s(\mu^2+1)-1\right]e^{s}\right] J_0(s\mu )\, \text{d}s,\nonumber\\
			&=\int_0^\infty \left[s\alpha(s)\left(\mu^2+1\right)-\beta(s)\right] J_0(s\mu )\, \text{d}s,
		\end{align}
		where we have defined the functions
		\begin{align}
			\alpha(s)=e^{-s}A(s)-e^{s}B(s),\quad \beta(s)=e^{-s}A(s)+e^{s}B(s).
		\end{align}	
		The right-hand side of these relations is reminiscent of the Hankel transform which is defined as
		\begin{align}
			\Hankeln{f(s)}=\int_{0}^{\infty}sf(s)J_n(s\mu )\, \text{d}s.
		\end{align}
		Moreover, $\Hankeln{\Hankeln{f}}=f$ and
		\begin{align}\label{eq:Hankelrel}
			\Hankel{\frac{e^{-s|\nu|}}{s}}=\frac{1}{\sqrt{\mu^2+\nu^2}},\quad	\Hankel{\frac{e^{-s|\nu|}}{|\nu|}}=\frac{1}{\left(\mu^2+\nu^2\right)^{3/2}}
		\end{align}
		However, in order to apply these relations to our boundary condition, we need to eliminate $\mu^2$ on the RHS of Eq.~\eqref{eq:BC2}. To this end we define the operator
		\begin{align}\label{eq:Lopdiff}
			\mathcal{L}=\frac{1}{s}\frac{d}{ds}\left[s\frac{d}{ds}\right]-1,
		\end{align}
		which satisfies
		\begin{align}
			\mathcal{L}\left[J_0(s\mu )\right]=-(\mu^2+1)J_0(s\mu ).
		\end{align}
		In particular, this allows us to write
		\begin{align}
			\int_{0}^{\infty}s\alpha(s)(\mu^2+1) J_0(s\mu )\, \text{d}s&=-\int_{0}^{\infty} s\alpha(s)\mathcal{L}\left[J_0(s\mu )\right]\, \text{d}s\nonumber\\
			&=-\int_{0}^{\infty}s\mathcal{L}\left[\alpha(s)\right]J_0(s\mu )\, \text{d}s,
		\end{align}
		where the second equality follows from two integrations by parts if we place the condition that $\alpha(0)$ is bounded and $\alpha\to0$ as $s\to\infty$. We may hence rewrite the boundary conditions to the diffusive problem as
		\begin{align}
			\frac{\nu_B+\xi}{\sqrt{\mu^2+\nu_B^2}}&=\int_0^\infty \left[\alpha(s)\sinh\left(s(1+\nu_B)\right)+\beta(s)\cosh\left(s(1+\nu_B)\right)\right] J_0(s\mu )\, \text{d}s,\\
			\frac{2\mathcal{Q}}{\sqrt{\mu^2+1}}&=\int_0^\infty \left[s\mathcal{L}\left[\alpha(s)\right]+\beta(s)\right] J_0(s\mu )\, \text{d}s,
		\end{align}
		and apply a Hankel transform to find
		\begin{align}
			\left(\nu_B+\xi\right)e^{-s\nu_B}&= \alpha(s)\sinh\left(s(1+\nu_B)\right)+\beta(s)\cosh\left(s(1+\nu_B)\right),\label{eq:bubbleBChankel}\\
			2\mathcal{Q}e^{-s}&=s\mathcal{L}\left[\alpha(s)\right]+\beta(s),
		\end{align}
		which gives
		\begin{align}
			\alpha''+\frac{1}{s}\alpha'-\left(1+\frac{\tanh\left(s(1+\nu_B)\right)}{s}\right)\alpha=\frac{2\mathcal{Q}}{s}e^{-s}-\frac{\nu_B+\xi}{s} e^{-s\nu_B}\sech\left(s(1+\nu_B)\right),\label{eq:diffODE} \\
			\beta=\left(\nu_B+\xi\right)e^{-s\nu_B}\sech\left(s(1+\nu_B)\right)-\tanh\left(s(1+\nu_B)\right)\alpha \label{eq:defbeta} ,
		\end{align}
		where primes denote derivatives with respect to $s$ and the boundary conditions are $\alpha(0)$ bounded and $\alpha\to0$ as $s\to\infty$. The ODE is singular at $s=0$, but this singularity is balanced by the forcing. Substituting a series solution of the form
		\begin{align}
			\alpha(s)=\sum_{n=0}^{\infty}\alpha_ns^n
		\end{align}
		readily shows that the correct lower boundary condition is $\alpha'(0)=\alpha_1=2\mathcal{Q}-\nu_B-\xi$. 
		
		Finally, we note that we can use Eqs.~\eqref{eq:nugrad} and \eqref{eq:Hankelrel} to find the flux integral
		\begin{align}\label{eq:fluxintegral}
			\int \hat{\bm{\nu}}\cdot\bm{\nabla}c\, \text{d}S&=\int_{0}^{2\pi}\int_{0}^{\infty} \frac{\partial c}{\partial \nu}\frac{h_\mu h_\theta}{h_\nu}\, \text{d}\mu\,\text{d}\theta=\begin{cases}
				8\pi\int_{0}^\infty A(s)\,\text{d}s, & \text{for } \nu >0 \\
				-8\pi\int_{0}^\infty B(s)\,\text{d}s, & \text{for } \nu < 0
			\end{cases},
		\end{align}
		which demonstrates that the flux out of the colloid (with $\nu<0$) is proportional to $\int_0^\infty B(s)\,\text{d}s$, while the flux into the bubble (with $\nu>0$) is proportional to $-\int_0^\infty A(s)\,\text{d}s$. Any residual between these is emitted to spatial infinity.
		
		\subsection{Example solutions}\label{sec:exsols}
		\paragraph{No bubble}\label{sec:nobubble}
		In the limit $\nu_B=\infty$, i.e\ a vanishing bubble, the system is radial symmetric about the centre of the colloid, and so we expect to recover the fundamental solution to Laplace's equation. The ODE for $\alpha$, Eq.~\eqref{eq:diffODE} reduces to
		\begin{align}
			\alpha''+\frac{1}{s}\alpha'-\left(\frac{1}{s}+1\right)\alpha=\frac{2\mathcal{Q}e^{-s}}{s},
		\end{align}
		and $\beta=-\alpha$. We note that this limit cannot be recovered smoothly, since there is always a boundary layer in the forcing for $s\lesssim1/(1+\nu_B)$ (corresponding to the contact region between bubble and colloid). This requires us to drop the boundary condition for $\alpha'(0)$ and just demand that the solution is bounded at $s=0$ and $\infty$. In that case the ODE has the particular integral
		\begin{align}
			\alpha(s)=-\mathcal{Q}e^{-s},\quad\beta(s)=\mathcal{Q}e^{-s},\\
			\Rightarrow A(s)=0,\quad B(s)=\mathcal{Q}e^{-2s}.
		\end{align}
		The two complementary functions for $\alpha$ are $e^s$ and $e^s\text{Ei}(-2s)$ which diverge at $s=\infty$ and $s=0$ respectively and are therefore discarded. Hence
		\begin{align}
			c-\tilde{c}^\infty&=(\mu^2+\nu^2)^{1/2}\int_0^\infty \mathcal{Q}e^{-s(2+\nu)}J_0(s\mu)\, \text{d}s\nonumber\\
			&=\mathcal{Q}\sqrt{\frac{\mu^2+\nu^2}{\mu^2+(\nu+2)^2}}\nonumber\\
			&=\frac{\mathcal{Q}}{\sqrt{\rho^2+(1+z)^2}},
		\end{align}
		as expected. From this we can conclude that the concentration scale on the colloid surface set by the flux condition is $\mathcal{Q}$. If Henry's law requires the concentration on a hypothetical bubble surface to be larger than this value, then the bubble will tend to dissipate. Furthermore, the flux integrals (see Eq.~\eqref{eq:fluxintegral}),
		\begin{align}
			\int_0^\infty A(s) \,\text{d}s= 0, \quad \int_0^\infty B(s) \,\text{d}s= \frac{\mathcal{Q}}{2},
		\end{align}
		inform us that there is no net flux into a surface of constant $\nu>0$ (as expected, since there is no bubble to absorb gas), and also that the flux out of the colloid in non-dimensional units is $4\pi\mathcal{Q}$, or $\mathcal{Q}$ per unit area. This again confirms expectations.
		
		\paragraph{A giant bubble}\label{sec:giantbubble}
		A giant bubble corresponds to the limit $\nu_B\to0$. We note that without making any assumptions about $\nu_B$, the boundary condition on the colloid, Eq.~\eqref{eq:defbeta}, implies that
		\begin{align}
			\beta&\approx 2\left(\nu_B+\xi\right)e^{-s(2\nu_B+1)}-\alpha\quad\text{for } s\gg\left(1+\nu_B\right)^{-1},\\
			\Rightarrow A(s)&\approx \left(\nu_B+\xi\right)e^{-2s\nu_B}\quad\text{for } s\gg\left(1+\nu_B\right)^{-1}.
		\end{align}
		Here, the range restriction on $s$ expresses the validity of the approximation only on the `far' side of the bubble. Due to our choice of length scaling, the range $s\lesssim\left(1+\nu_B\right)^{-1}$ actually describes a boundary layer wherein the colloid touches the bubble and perturbs the flux of solute locally. It is therefore more instructive to reinsert dimensions and instead take the limit of a vanishing colloid, $R_c\to0$. It may then be seen that the boundary condition on the colloid, Eq.~\eqref{eq:dbc2app}, reduces to the condition $B(s)=0$, whereupon the asymptotic for $A(s)$ becomes exact and extends to the whole range. Introducing the scaled variable $w=(R_c/R_B)s$ we have (in dimensional units)
		\begin{align}
			A(w)\approx\frac{2\gamma}{K_H}\left(\frac{1}{R_B}+\xi^*\right)e^{-2w},\quad B(w)\approx0\qquad\text{for } w\gg\frac{R_c}{R_B+R_c},
		\end{align}
		where $\xi^*=\xi/R_c$ remains finite when the limit $R_c\to 0$ is taken. In analogy to the case of no bubble, we again obtain the fundamental solution to Laplace's equation for the concentration field,
		\begin{align}
			c-c^\infty&=\frac{2\gamma}{K_H}\left(\frac{1}{R_B}+\xi^*\right)\sqrt{\frac{\mu^2+\nu^2}{\mu^2+(\nu-2\nu_B)^2}}=\frac{2\gamma}{K_H}\frac{1+R_B\xi^*}{\sqrt{\rho^2+(z-R_B)^2}}.
		\end{align}
		We note that the net catalytic flux out of the bubble scales as $2\gamma D(1+\xi^*R_B)/K_H$. It therefore grows without bound with the bubble size unless the background fluid is completely saturated ($\xi^*=0$). Hence, unless this condition is met, there is an upper limit to bubble growth for any catalytic flux $\mathcal{Q}$.

		\subsection{First order kinetics}\label{sec:1ok}
		
		\paragraph{Model and method of solution}
		In this section we examine first-order kinetics, which are a more realistic model of the chemical reaction on the colloid. Instead of assuming production of a constant flux of reaction product (oxygen), $\mathcal{A}$, we instead assume the presence of a second, `fuel' field (hydrogen peroxide), $c_f$, that is converted on the colloid surface at a constant rate, $k$. This leads to two boundary conditions of the form
		\begin{equation}\label{eq:diffbc2_1ok}
			\begin{rcases}
				-D\hat{\bm{\nu}}\cdot\bm{\nabla}c &= k c_f\\
				-D_f\hat{\bm{\nu}}\cdot\bm{\nabla}c_f &= -k c_f
			\end{rcases}\quad\text{at }\nu=-\nu_c\quad\textrm{(1ok)}.
		\end{equation}
		It is additionally necessary to solve diffusive dynamics for the fuel field, $D_f\nabla^2c_f=0$, and specify another boundary condition on the bubble. Since hydrogen peroxide is well soluble in water, we can assume that the bubble consists exclusively of reaction product, i.e.~that there is no fuel flux in or out of the bubble. Then
			\begin{align}\label{eq:diffbc1_fuel}
				-D_f\hat{\bm{\nu}}\cdot\bm{\nabla}c_f=0,\quad\text{at }\nu=\nu_B.
			\end{align}
		Finally, we assume that far away the fuel concentration asymptotes to some bulk value $c_f^\infty$ that supplies the reaction.
		
		Using the same scalings as before, the four boundary conditions on bubble and colloid then become
		\begin{align}
			\frac{\partial c_f}{\partial \nu}=0,&\qquad c-\tilde{c}^\infty=\xi+\nu_B \quad\text{at }\nu=\nu_B,\\
			\frac{\partial c_f}{\partial \nu} = \frac{2\,\Da}{\mu^2+1}c_f,&\qquad \frac{\partial c}{\partial \nu} = -\frac{2\,\Delta\,\Da}{\mu^2+1}c_f \quad\text{at }\nu=-1,
		\end{align}
		and feature two new non-dimensional parameters,
		\begin{align}
			\Da=\frac{kR_c}{D_f},\quad \Delta=\frac{D_f}{D},
		\end{align}
		The Damk\"ohler number $\Da$ expresses the ratio of reaction speed to the diffusive recruitment of reaction fuel. A large Damk\"ohler number indicates that fuel is depleted faster than it is replenished, which may be expected to inhibit bubble growth. The parameter $\Delta$ is just the ratio of diffusivities of product and fuel and not very important for the dynamics.
		
		We note that if we define another parameter $\tilde{\mathcal{Q}}$ analogously to $\mathcal{Q}$ in zeroth order kinetics, with $\mathcal{A}$ replaced by $kc_f^\infty$, we can write the product boundary condition on the colloid as
		\begin{align}
			\frac{\partial c}{\partial \nu}  = -\frac{2\tilde{\mathcal{Q}}}{\mu^2+1}\frac{c_f}{\tilde{c}_f^\infty}\quad\text{at }\nu=-1,
		\end{align}
		which is more reminiscent of its form for zeroth order kinetics, Eq.~\eqref{eq:dbc2app}. This limit then corresponds formally to $\Da\to 0$ with $\tilde{\mathcal{Q}}=\mathcal{Q}$ constant (and $\tilde{c}_f^\infty=\mathcal{Q}/\left(\Delta\,\Da\right) \to\infty$), i.e.~a slow reaction with a large fuel supply. In summary, first order kinetics are determined by four dimensionless parameters, $\{\tilde{\mathcal{Q}},\xi,\text{Da},\Delta\}$, that generalise the two dimensionless parameters of zeroth order kinetics.
		
		To solve the problem, we use the solution ansatz
		\begin{align}
			c_f(\mu,\nu) &= c_f^\infty+\sqrt{\mu^2+\nu^2} \int_{0}^\infty \left[ c_f^\infty A_f(s)e^{s\nu}+ c_f^\infty B_f(s)e^{-s\nu}\right]J_0(s\mu)\,\text{d}s,\\
			c(\mu,\nu) &= c^\infty+\sqrt{\mu^2+\nu^2} \int_{0}^\infty \left[  A(s)e^{s\nu}+ B(s)e^{-s\nu}\right]J_0(s\mu)\,\text{d}s
		\end{align}
		which satisfies Laplace's equation and the boundary conditions at spatial infinity by construction. We note that here $A_f$ and $B_f$ have been defined with a prefactor that leads to more elegant dimensionless expressions. Substituting, rearranging and Hankel-transforming as before leads to four coupled second order differential equations,
		\begin{align}
			\alpha_f''+\frac{\alpha_f'}{s}&-\left(\nu_B^2+\frac{\nu_B \coth(1+\nu_B)s}{s}\right)\alpha_f -\frac{\nu_B}{s \sinh(1+\nu_B)s}\beta_f=0,\\
			\beta_f''+\frac{\beta_f'}{s}&-\left(1+\frac{(1+2\,\Da)\coth(1+\nu_B)s}{s}\right)\beta_f-\frac{1+2\,\Da}{s \sinh(1+\nu_B)s}\alpha_f= \frac{2 \,\Da }{s}e^{-s},\\
			\alpha''+\frac{\alpha}{s}'&-\left(1+\frac{\tanh\left(s(1+\nu_B)\right)}{s}\right)\alpha-\frac{2\tilde{\mathcal{Q}}}{s}\beta_f=\frac{2\tilde{\mathcal{Q}}}{s}e^{-s}-\frac{\nu_B+\xi}{s} e^{-s\nu_B}\sech\left(s(1+\nu_B)\right),\\
			\beta&=\left(\nu_B+\xi\right)e^{-s\nu_B}\sech\left(s(1+\nu_B)\right)-\tanh\left(s(1+\nu_B)\right)\alpha,
		\end{align}
		where $\alpha_f$ and $\beta_f$ are defined as
		\begin{align}
			\alpha_f (s)&= e^{\nu_B s}A_f(s)-e^{-\nu_B s}B_f(s)\\
			\beta_f (s)&= -e^{-s}A_f(s)+e^{s}B_f(s).
		\end{align}
		The first two of these equations uncouple from the rest and we can solve them numerically to find $\alpha_f$ and $\beta_f$. The boundary conditions are such that both $\alpha_f$ and $\beta_f$ decay to zero at infinity (from the integration by parts), and that  $\alpha_f'(0)=-\nu_B$ and $\beta_f'(0)=-1$ (from requiring regularity of the solution at $s=0$). Then with $\beta_f$ in hand we can solve the second pair of equations to find $\alpha$ and $\beta$, with $\alpha'(0)=2\tilde{\mathcal{Q}}\left(1+\beta_f(0)\right)-\nu_B-\xi$ and $\alpha\to 0$ as $s\to\infty$. Both of these steps are straightforward using a standard boundary value problem solver. This completes the solution. The case of zeroth-order kinetics is smoothly recovered as $\Da\to 0$, in which case $\alpha_f,\beta_f\to 0$, and the equations and boundary conditions for the product field reduce appropriately.
		
		\paragraph{Interpretation}
		
		\begin{figure}[t]
			\centering
			\begin{subfigure}{0.35\textwidth}
				\includegraphics[width=\textwidth]{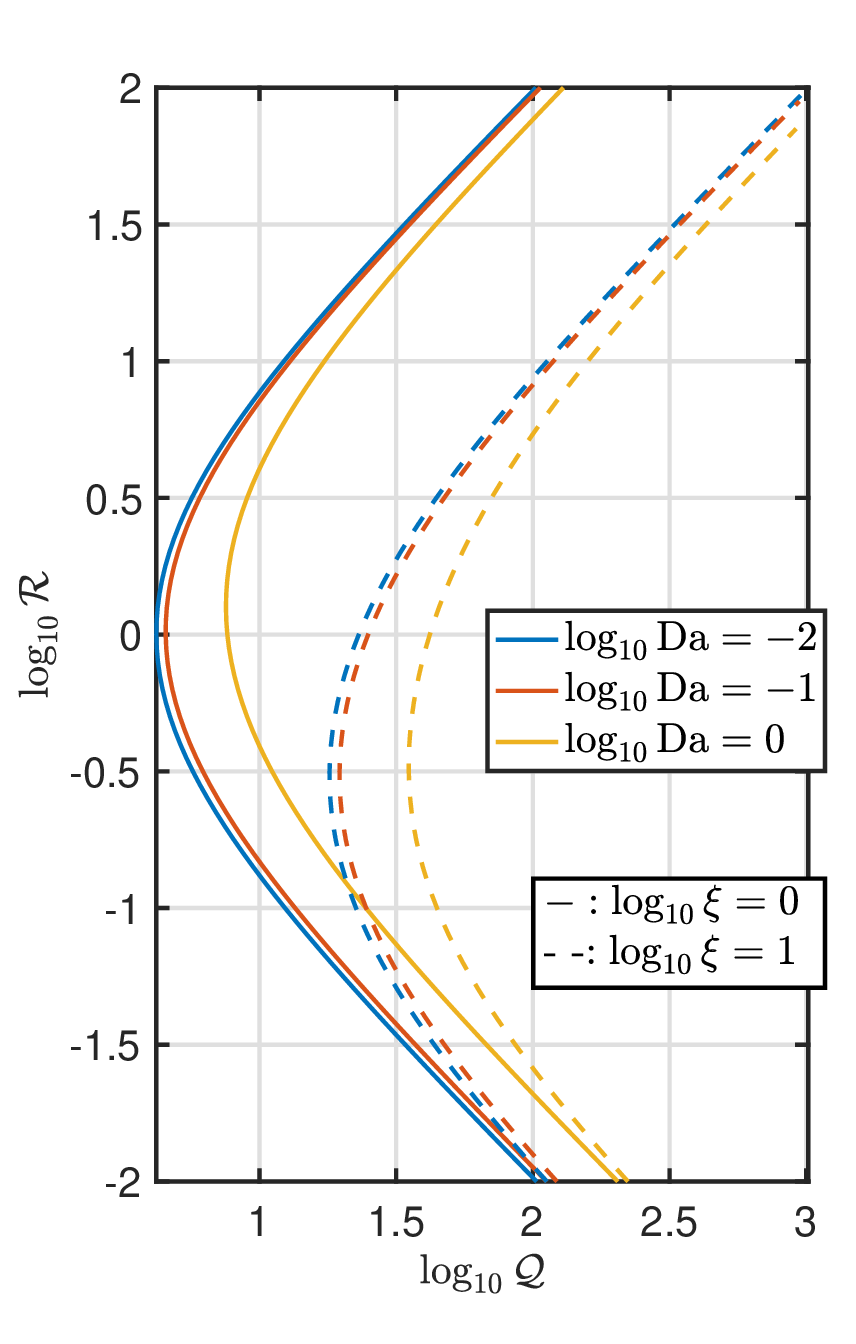}
			\end{subfigure}
			\hfill
			\begin{subfigure}{0.64\textwidth}
				\includegraphics[width=\textwidth]{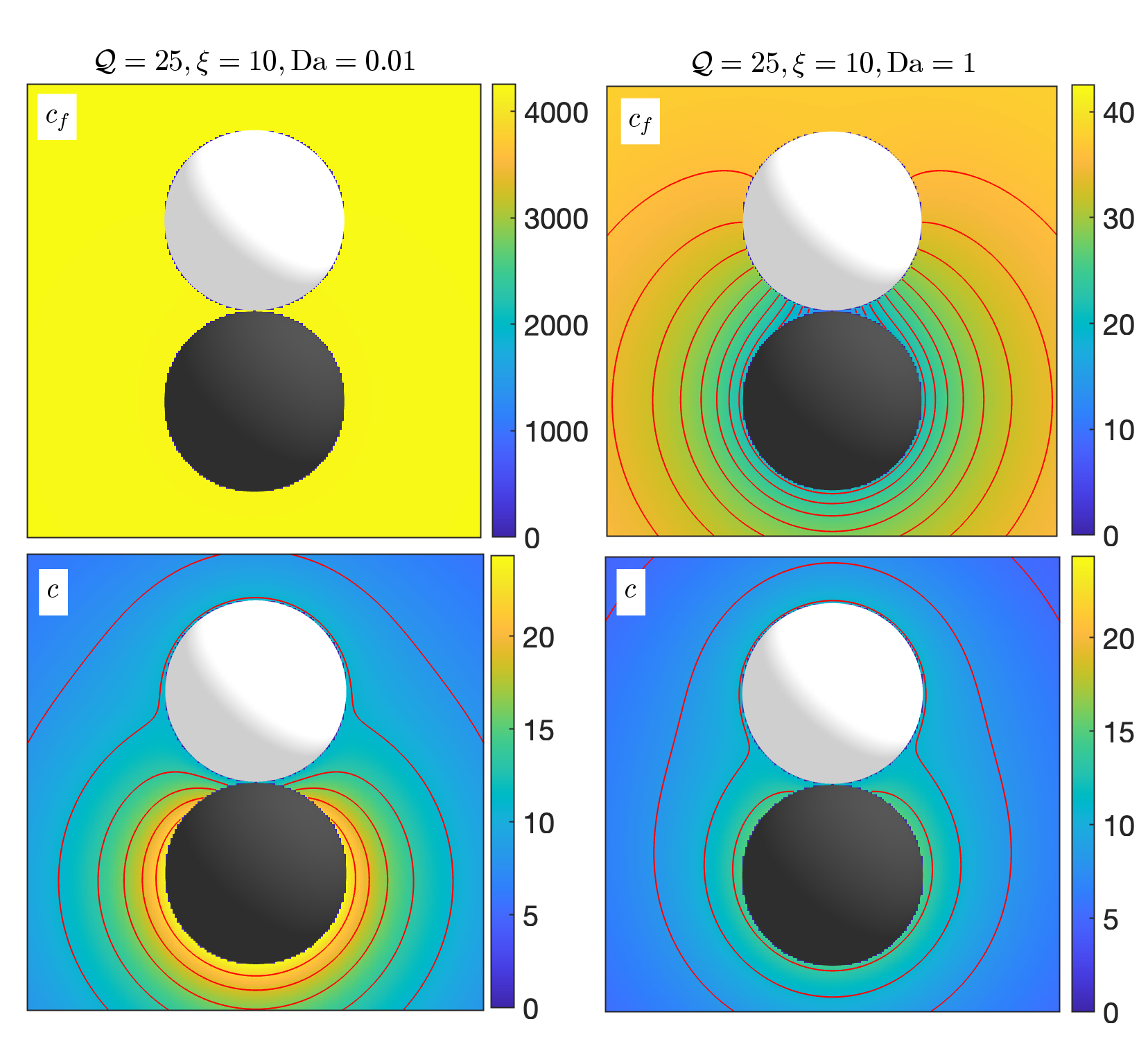}
			\end{subfigure}
			\caption{Effect of the Damk\"ohler number Da on the bubble growth dynamics. Left: Dashed and dash-dotted lines indicate theoretical estimates for $R_\text{max}$ (stable) and $R_\text{min}$ (unstable) respectively. The effect of a non-zero Damk\"ohler number is approximately a shift of the nullclines towards larger values of flux $\mathcal{Q}$. Right: Illustration of the fuel (top) and solute (bottom) concentration fields for small and large Da.}\label{fig:1ok}
		\end{figure}
		
		For typical values of the parameters (see section \ref{sec:paramestimates}), we can estimate the new parameters as
		\begin{align}
			\Da=2.9 R_c \times \SI{e-2}{\per\micro\metre},\quad \Delta=0.59,
		\end{align}
		and so the Damk\"ohler number typically ranges from $\mathcal{O}(10^{-1})-\mathcal{O}(1)$. We illustrate the effect of Da in Fig.~\ref{fig:1ok}. For small values, the concentration fields are essentially indistinguishable from zeroth-order kinetics, as one might expected. However, even for Da of order unity, the effect of including first-order kinetics is mostly an overall reduction of the flux $\mathcal{Q}$ out of the colloid, resulting in a shift of the line of stationary points in the phase diagram. In particular, there is no excessive perturbation to the $c$ field in the gap between bubble and colloid, despite the depletion of fuel there. As a result, the time-dependent dynamics described in the main text remain unchanged, save for a quantitative shift in the colloid activity $\mathcal{Q}$ that is required to achieve a certain bubble size.

	\section{Detailed solution of the quasi-steady fluid mechanical model}\label{sec:app_fluid}
	
	\subsection{Setup}
	The classical work on Stokes flows in this geometry was carried out in Ref.~\cite{cooley1969slow} for the motion of two rigid spheres in contact, and generalised in Ref.~\cite{reed1974slow} to fluid spheres. Here we broadly follow Ref.~\cite{cooley1969slow} in developing the solution ansatz. Since the flow is axisymmetric, we employ a stream function to solve the problem. In cylindrical coordinates $\{\rho,\theta,z\}$, such a function $\Psi$ is defined by
	\begin{align}
		u_\rho=\rho^{-1}\partial_z\Psi, \quad u_z=-\rho^{-1}\partial_\rho\Psi.
	\end{align}
	It may then be shown that this flow satisfies
	\begin{align}\label{eq:munustrfun}
		\bm{u}=\frac{\left(\mu^2+\nu^2\right)^2}{4\mu}\left(-\frac{\partial\Psi}{\partial\nu}\hat{\bm{\mu}}+\frac{\partial\Psi}{\partial\mu}\hat{\bm{\nu}}\right)
	\end{align}
	in tangent sphere coordinates. Then, it can be shown that the Stokes equations reduce to
	\begin{align}
		\Lambda^4\Psi=0,
	\end{align}
	where the Stokes operator $\Lambda^2$ is given by
	\begin{align}
		\Lambda^2&=\rho\partial_\rho\left(\rho^{-1}\partial_\rho\right)+\partial_{zz}\\
		&=\frac{1}{4}\mu(\mu^2+\nu^2)\left[\partial_\mu\left(\mu^{-1}(\mu^2+\nu^2)\partial_\mu\right)+\partial_\nu\left(\mu^{-1}(\mu^2+\nu^2)\partial_\nu\right)\right].
	\end{align}
	A solution to the Stokes equations $\Psi$ may be constructed from solutions to $\Lambda^2\phi=0$ by adding linearly independent resonant terms. By considering the form of the $\Lambda^2$ operator in spherical and cylindrical coordinates respectively it may be shown that one way to achieve this is by writing
	\begin{align}\label{eq:solconstr}
		\Psi=(\rho^2+z^2)\phi+z\chi,
	\end{align}
	where $\Lambda^2\phi=\Lambda^2\chi=0$. In this case we have irrotational flow, $\Lambda^2\Psi=0$, if and only if
	\begin{equation}
		\phi+2\rho\phi_\rho+2z\phi_z+\chi_z=0.
	\end{equation}
	The reason we choose this representation among several equivalent ones (such as $\Psi=\phi+z\chi$) is due to it leading to a convenient form of the stream function later on.
	
	In order to solve $\Lambda^2\phi=0$ we write
	\begin{align}
		\phi=\frac{\mu}{\sqrt{\mu^2+\nu^2}}F[\mu,\nu].
	\end{align}
	It may then be shown~\cite{cooley1969slow} that
	\begin{align}
		\Lambda^2\phi=0\quad\Leftrightarrow\quad \frac{1}{4\mu}(\mu^2+\nu^2)^{3/2}\left(\mu^2F_{\mu\mu}+\mu F_\mu+\mu^2 F_{\nu\nu}-F\right)=0.
	\end{align}
	and with the ansatz $F=\exp(\pm s\nu)f(s\mu)$ we see that $f$ satisfies Bessel's equation of order one. The solution bounded for $\mu\to 0$ (i.e.~no radial flow on the $z$-axis) may hence be written as
	\begin{align}
		F[\mu,\nu]=\int_0^\infty \left[f_+(s)e^{s\nu}+f_-(s)e^{-s\nu}\right]J_1(s\mu)\,\text{d}s,
	\end{align}
	where the functions $f_+$ and $f_-$ are determined by the boundary conditions. Using the definition of the coordinates $\mu$ and $\nu$ in Eq.~\eqref{eq:solconstr} we then have a general solution of the form
	\begin{align}
		\Psi=\frac{\mu}{\left(\mu^2+\nu^2\right)^{3/2}}\int_0^\infty\left[\left(A+C\nu\right)\sinh{s\nu}+\left(B+D\nu\right)\cosh{s\nu}\right]J_1(s\mu)\,\text{d}s,
	\end{align}
	where the coefficients $A$ to $D$ are functions of $s$ and to be determined from the boundary conditions. We stress that these are different from the coefficients of the diffusive problem, but in this section we drop the notation with a $\sim$ used in the main text to avoid clutter.  We note however, that since $J_1(s\mu)\to0$ as $\mu\to 0$, this ansatz is only capable of describing flows that have $\Psi\equiv0$ on the axis of symmetry. This is the case for pure translation, but not for bubble growth since volume is not conserved. For this reason it is necessary to add a singular correction $\Psi_s$ that satisfies $\Psi_s=2\dot{R}_B/\nu_B^2$ for $z\geq2R_B$ and $\Psi_s=0$ for $z\leq0$ on the axis of symmetry. This is achieved by adding the stream function of a point source at $z=R_B$~\cite{michelin2018collective}. In terms of $\{\mu,\nu\}$ coordinates this is
	\begin{align}
		\Psi_s = \frac{\dot{R}_B}{\nu_B^2}\left(1+\frac{2\nu\nu_B-\mu^2-\nu^2}{\sqrt{\left(\mu^2+\nu^2\right)\left(\mu^2+\left(\nu-2\nu_B\right)^2\right)}}\right).
	\end{align}
	The complete stream function is hence
	\begin{align}\label{eq:genstokes}
		\Psi=\Psi_s+\frac{\mu}{\left(\mu^2+\nu^2\right)^{3/2}}\int_0^\infty\left[\left(A+C\nu\right)\sinh{s\nu}+\left(B+D\nu\right)\cosh{s\nu}\right]J_1(s\mu)\,\text{d}s.
	\end{align}
	From Ref.~\cite{happel2012low} we have the formula 4-14.18 for the force on the body in the $z$ direction,
	\begin{align}
		F_z=\pi\eta R_c\int \rho^3 \frac{\partial}{\partial n}\left(\frac{\Lambda^2 \Psi}{\rho^2}\right)\,\text{d}\iota,
	\end{align}
	where $n$ is the outward normal, $\eta$ is the fluid viscosity and d$\iota$ is the tangential length increment with a negative $z$-component (see figure 4-5.1 in Ref.~\cite{happel2012low}). Thus,
	\begin{align}
		\text{d}\iota \frac{\partial}{\partial n} = -\frac{h_\mu}{h_\nu}\text{d}\mu \frac{\partial}{\partial \nu}=-\text{d}\mu \frac{\partial}{\partial \nu}
	\end{align}
	both on the top and on the bottom sphere. We then find the forces on bubble, colloid, and the dimer to be
	\begin{align}\label{eq:forceintegral1}
		F_z^{bub} &= \pi\eta R_c \int_0^\infty s(B(s)+A(s))\,\text{d}s,\\
		F_z^{col} &= \pi\eta R_c \int_0^\infty s(B(s)-A(s))\,\text{d}s,\label{eq:forceintegral2}\\
		F_z^{tot} &= 2\pi\eta R_c \int_0^\infty sB(s)\,\text{d}s.\label{eq:forceintegral}
	\end{align}
	The final result can alternatively be obtained using the limit formula (4-14.19 in Ref.~\cite{happel2012low}),
	\begin{align}
		F_z^{tot}=8\pi\eta R_c\lim_{r\to\infty}\frac{r\Psi}{\rho^2},
	\end{align}
	where $r=\sqrt{\rho^2+z^2}$.
	
	\subsection{Solution for no slip conditions on the bubble}
	Suppose the bubble-colloid system is moving with velocity $U$ in the negative $z$-direction (i.e.~with the colloid at the front), and the radius $R_B$ of the bubble grows at a rate $\dot{R}_B$. Then in the lab frame, the flow has velocity zero at infinity, $-U\hat{\bm{z}}$ on the colloid, and on the bubble it is $\dot{R}_B(\hat{\bm{z}}-\hat{\bm{\nu}})-U\hat{\bm{z}}$. Using Eq.~\eqref{eq:coordconv} , the velocity on the bubble $\nu=\nu_B$ is found to be
	\begin{align}
		\bm{u}=\left(U-\dot{R}_B\right)\frac{2\mu\nu_B}{\mu^2+\nu_B^2} \hat{\bm{\mu}}-\left(\left(U-\dot{R}_B\right)\frac{\mu^2-\nu_B^2}{\mu^2+\nu_B^2}
		+\dot{R}_B\right)\hat{\bm{\nu}},
	\end{align}
	while on the colloid at $\nu=-1$ it is
	\begin{align}
		\bm{u}=-U\frac{2\mu}{\mu^2+1}\hat{\bm{\mu}} -U\frac{\mu^2-1}{\mu^2+1} \hat{\bm{\nu}}.
	\end{align}
	By comparing with the expression for the stream function in these coordinates, Eq.~\eqref{eq:munustrfun}, we find the boundary conditions
	\begin{align}
		\Psi\vert_{\nu=-1}&=U\frac{2\mu^2}{\left(\mu^2+1\right)^2},\\
		\frac{\partial\Psi}{\partial\nu}\bigg\vert_{\nu=-1}&=U\frac{8\mu^2}{\left(\mu^2+1\right)^3},\\
		\Psi\vert_{\nu=\nu_B}&=\left(U-\dot{R}_B\right)\frac{2\mu^2}{\left(\mu^2+\nu_B^2\right)^2}+\dot{R}_B\frac{2}{\mu^2+\nu_B^2},\label{eq:bubnopen}\\
		\frac{\partial\Psi}{\partial\nu}\bigg\vert_{\nu=\nu_B}&=-\left(U-\dot{R}_B\right)\frac{8\mu^2\nu_B}{\left(\mu^2+\nu_B^2\right)^3},
	\end{align}
	where a constant of integration has been chosen so that $\Psi\to0$ as $\mu\to\infty$ and $\mu\to 0$ with $\nu<0$, and $\Psi\to2\dot{R}_B/\nu_B^2$ as $\mu\to 0$ with $0<\nu<\nu_B$, thus matching the behaviour of $\Psi_s$. We solve for the motility and growth problems separately. 
	
	Writing $\tilde{\Psi}=\Psi-\Psi_s$, we note that the $\nu$-derivative of the streamfunction is related to $\tilde{\Psi}$ by
	\begin{align}
		\frac{\partial\tilde{\Psi}}{\partial\nu}=\frac{-3\nu}{\mu^2+\nu^2}\tilde{\Psi} +\frac{\mu}{\left(\mu^2+\nu^2\right)^{3/2}}\int_0^\infty\left\{\left[C+\left(B+D\nu\right)s\right]\sinh{s\nu}+\left[D+\left(A+C\nu\right)s\right]\cosh{s\nu}\right\}J_1(s\mu)\,\text{d}s.
	\end{align}
	Hence,
	\begin{align}
		\left[Cs^{-1}+\left(B+D\nu\right)\right]\sinh{s\nu}+\left[Ds^{-1}+\left(A+C\nu\right)\right]\cosh{s\nu} = \mathcal{H}_1\left[\frac{\left(\mu^2+\nu^2\right)^{3/2}}{\mu}\left(\frac{\partial\tilde{\Psi}}{\partial\nu}+\frac{3\nu}{\mu^2+\nu^2}\tilde{\Psi} \right) \right].
	\end{align}
	At the same time, it follows from a direct Hankel transform of $\Psi$ that
	\begin{align}
		\left(A+C\nu\right)s^{-1}\sinh{s\nu}+\left(B+D\nu\right)s^{-1}\cosh{s\nu} =  \mathcal{H}_1\left[\frac{\left(\mu^2+\nu^2\right)^{3/2}}{\mu}\tilde{\Psi}\right].
	\end{align}
	Evaluated at $\nu=\{\nu_B,-1\}$ these two relations give an algebraic system of four equations that determines the four coefficients in the solution. For the motility problem,
	\begin{align}
		-\left(A-C\right)s^{-1}\sinh{s}+\left(B-D\right)s^{-1}\cosh{s} &= 2U \left(s^{-2}+s^{-1}\right)e^{-s},\\
		-\left[Cs^{-1}+\left(B-D\right)\right]\sinh{s}+\left[Ds^{-1}+\left(A-C\right)\right]\cosh{s} &=  2Ue^{-s},\\
		\left(A+C\nu_B\right)s^{-1}\sinh{s\nu_B}+\left(B+D\nu_B\right)s^{-1}\cosh{s\nu_B} &= 2U \left(s^{-2}+\nu_Bs^{-1}\right)e^{-s\nu_B},\\
		\left[Cs^{-1}+\left(B+D\nu_B\right)\right]\sinh{s\nu_B}+\left[Ds^{-1}+\left(A+C\nu_B\right)\right]\cosh{s\nu_B} &= -2 U \nu_B e^{-s\nu_B},
	\end{align}
	while for the growth problem the right hand side of these equations reads
	\begin{align}
		\left\{\dot{R}_B\nu_B^{-2}s^{-4}\left(-3(1+s)+3e^{2\nu_B s}(1+s)-2\nu_B s(3+s(3+4\nu_B(1+s+\nu_B s)))\right)e^{-(1+2\nu_B)s} ,\right.\nonumber\\
		\dot{R}_B\nu_B^{-2}s^{-2}\left(-3+3e^{2\nu_B s}-2\nu_B s (3+4\nu_B(1+\nu_B)s)\right)e^{-(1+2\nu_B)s} ,\nonumber\\
		-2\dot{R}_Bs^{-2}\left(1+\nu_B s\right)e^{-\nu_Bs} ,\nonumber\\
		\left.2\dot{R}_B\nu_B e^{-\nu_Bs}\right\}
	\end{align}
	The solution to this is quite messy, and since we only need the coefficient $B$ for the force calculation, we will only report that one here. For the motility problem,
	\begin{align}
		B_\text{mot}=2 U\times & \bigg[\left(2 s^2+2 s+1\right) w +\left(2 \nu_B^2 s^2+2 \nu_B s+1\right)
		w^{\nu_B}-\left(2 \nu_B^2 s^2-2 \nu_B s+1\right) w^{\nu_B+2}\nonumber\\
		&-\left(8 \nu_B (\nu_B+1) s^3+4 (\nu_B+1)^2 s^2+4 (\nu_B+1)
		s+2\right) w^{\nu_B+1}\nonumber\\
		&-\left(2 s^2-2 s+1\right) w^{2 \nu_B+1}+2 w^{2 \nu_B+2}\bigg]\bigg{/}\nonumber\\
		&\big[ s \left(1-\left(4 (\nu_B+1)^2 s^2+2\right)
		w^{\nu_B+1}+w^{2 \nu_B+2}\right)\big],
	\end{align}
	while for the growth problem,
	\begin{align}
		B_\text{gro}=-\dot{R}\times&\bigg[2 \nu_B^2 s^2 (2 \nu_B s+1) w^{\nu_B}+2 \left(2 \nu_B^2 (2 \nu_B+3) s^4+2 \nu_B \left(\nu_B^2+3 \nu_B+3\right) s^3\right.\nonumber\\
		&\left.+3 \left(\nu_B^2+4
		\nu_B+2\right) s^2+6 (\nu_B+1) s+3\right) w^{\nu_B+1}-\left(4 \nu_B^2 \left(8 \nu_B^2+12 \nu_B+3\right) s^4\right.\nonumber\\
		&\left.+4 \nu_B \left(6 \nu_B^2+11
		\nu_B+3\right) s^3+6 \left(3 \nu_B^2+4 \nu_B+1\right) s^2+6 (2 \nu_B+1) s+3\right) w^{2 \nu_B+1}\nonumber\\
		&+\left(-8 \nu_B^3 s^3-6 \nu_B^2 s^2+6\right) w^{2
			\nu_B+2}-\left(8 \nu_B^3 s^3+8 \nu_B^2 s^2+6 \nu_B s+3\right) w^{3 \nu_B+2}\nonumber\\
		&+(6 \nu_B s-3) w^{\nu_B+2}-3 \left(2 s^2+2 s+1\right) w\bigg]\bigg{/}\nonumber\\
		&\big[\nu_B^2 s^3 \left(1-\left(4 (\nu_B+1)^2 s^2+2\right)
		w^{\nu_B+1}+w^{2 \nu_B+2}\right)\big],
	\end{align}
	where $w=e^{-2s}$. As $s\to 0$, the coefficient $A_\text{gro}\sim -s^{-2}\to-\infty$ diverges, which according to Eqs.~\eqref{eq:forceintegral1}-\eqref{eq:forceintegral2} leads to an infinite attractive force between the two spheres. The regularisation of this force is discussed in section \ref{sec:pressure}. Unlike for the diffusive problem, there is no need to solve an ODE since $\Psi$ and $\partial\Psi/\partial\nu$ are specified on the same surfaces. This is no longer the case if we replace the boundary condition on the bubble with no shear stress.
	
	
	\subsection{Solution for no stress condition on the bubble}
	
	In this section we consider an alternative boundary condition on the bubble $\nu=\nu_B$. As in the rigid case, we have a no penetration condition for the normal $\hat{\bm{\nu}}$-component of the velocity. In terms of the stream function, this reads (c.f. Eq.~\eqref{eq:bubnopen}).
	\begin{align}\label{eq:nopenalt}
		\Psi\vert_{\nu=\nu_B} &=U\frac{2\mu^2}{\left(\mu^2+\nu_B^2\right)^2}+\dot{R}_B\frac{2\nu_B^2}{\left(\mu^2+\nu_B^2\right)^2}.
	\end{align}
	The condition for the tangential $\hat{\bm{\mu}}$-component is replaced by a no stress condition, $\hat{\bm{\nu}}\cdot\left(\nabla\bm{u}+\left(\nabla\bm{u}\right)^T\right)\cdot\hat{\bm{\mu}}=0$. With some algebra it may be shown that this is equivalent to		
	\begin{align}\label{eq:noshearbcstep}
		\left[\Lambda^2\Psi+\nu_B\left(\mu^2+\nu_B^2\right)\frac{\partial\Psi}{\partial\nu}+3\nu_B^2\Psi\right]_{\nu=\nu_B} &=\left[2\mu\frac{\partial}{\partial \mu}\left(\frac{\left(\mu^2+\nu_B^2\right)^2}{4\mu }\frac{\partial\Psi}{\partial \mu}\right)+3\nu_B^2\Psi\right]_{\nu=\nu_B}.
	\end{align}
	The boundary conditions on the colloid remain unchanged.  Using the no penetration condition Eq.~\eqref{eq:nopenalt} to simplify the right-hand side of Eq.~\eqref{eq:noshearbcstep}, we arrive at
	\begin{align}
		\Psi\vert_{\nu=-1}&=U\frac{2\mu^2}{\left(\mu^2+1\right)^2},\label{eq:BChyd1}\\
		\frac{\partial\Psi}{\partial\nu}\bigg\vert_{\nu=-1}&=U\frac{8\mu^2}{\left(\mu^2+1\right)^3},\label{eq:BChyd2}\\
		\Psi\vert_{\nu=\nu_B}&=U\frac{2\mu^2}{\left(\mu^2+\nu_B^2\right)^2}+\dot{R}_B\frac{2\nu_B^2}{\left(\mu^2+\nu_B^2\right)^2},\label{eq:BChyd3}\\
		\left[\Lambda^2\Psi+\nu_B\left(\mu^2+\nu_B^2\right)\frac{\partial\Psi}{\partial\nu}+3\nu_B^2\Psi\right]_{\nu=\nu_B} &=-U\frac{2\mu^2\nu_B^2}{\left(\mu^2+\nu_B^2\right)^2}+\dot{R}_B\frac{8\mu^2\nu_B^2+6\nu_B^4}{\left(\mu^2+\nu_B^2\right)^2}\label{eq:BCnostress}
	\end{align}
	
	With the boundary conditions posed in this form, it turns out that there is an elegant path to a fully analytical solution. First, note that by performing two Hankel transforms of different orders we have the following identity,
	\begin{align}
		\left(A+C\nu_B\right)\sinh{s\nu_B}+\left(B+D\nu_B\right)\cosh{s\nu_B}
		=&\, s \mathcal{H}_1\left[\frac{\left(\mu^2+\nu_B^2\right)^{3/2}}{\mu}\tilde\Psi\vert_{\nu=\nu_B} \right]\nonumber\\
		=&\,2\left(U-\dot{R}_B\right) \left(s^{-1}+\nu_B\right)e^{-s\nu_B}\nonumber\\
		=&\,\left(U-\dot{R}_B\right) \times \mathcal{H}_0\left[\frac{2(\mu^2+2\nu_B^2)}{(\mu^2+\nu_B^2)^{3/2}} \right].\label{eq:HankelIdentity}
	\end{align}
	Furthermore, from differentiating the general form of the streamfunction Eq.~\eqref{eq:genstokes} we have that
	\begin{align}
		&\left[\Lambda^2\tilde\Psi+\nu_B\left(\mu^2+\nu_B^2\right)\frac{\partial\tilde\Psi}{\partial\nu}+3\nu_B^2\tilde\Psi\right]_{\nu=\nu_B}\nonumber\\
		=&\,  -\frac{\mu^2}{(\mu^2+\nu_B^2)^{1/2}} \mathcal{H}_0\left[\left(A+C\nu_B\right)\sinh{s\nu_B}+\left(B+D\nu_B\right)\cosh{s\nu_B}\right]\nonumber\\
		&\, +\frac{1}{2}\frac{\mu}{(\mu^2+\nu_B^2)^{1/2}}\int_{0}^{\infty} s\left[3\left(\beta+\nu_B\delta\right)+(\mu^2+\nu_B^2)\gamma \right]J_1(s\mu)\,\text{d}s,
	\end{align}
	where we define Greek coefficients as
	\begin{align}
		\alpha(s) =  \, A \cosh(s\nu_B) +B  \sinh(s\nu_B), \qquad& \gamma(s) =     C \cosh(s\nu_B)+D  \sinh(s\nu_B),\nonumber\\
		\beta(s) =  \, s^{-1}\left[B  \cosh(s\nu_B)+  A \sinh(s\nu_B)\right],\qquad &\delta(s) =  s^{-1}\left[D  \cosh(s\nu_B)+  C \sinh(s\nu_B)\right],
	\end{align}
	($\alpha$ is defined only for completeness). After substituting the no stress condition Eq.~\eqref{eq:BCnostress} and applying the identity Eq.~\eqref{eq:HankelIdentity} this simplifies to
	\begin{align}\label{eq:hydintermediatestep}
		\int_{0}^{\infty} s\left[3\left(\beta+\nu_B\delta\right)+(\mu^2+\nu_B^2)\gamma  \right]J_1(s\mu)\,\text{d}s =  \left(U-\dot{R}_B\right)\frac{4\mu}{\left(\mu^2+\nu_B^2\right)^{1/2}}.
	\end{align}
	In analogy with the diffusive case (c.f.\ Eq.~\eqref{eq:Lopdiff}), we define the self-adjoint operator
	\begin{align}
		\bar{\mathcal{L}}=\frac{1}{s}\frac{\text{d}}{\text{d}s}\left[s\frac{\text{d}}{\text{d}s}\right]-\frac{1}{s^2}-\nu_B^2,
	\end{align}
	which satisfies
	\begin{align}
		\bar{\mathcal{L}}\left[J_1(s\mu )\right]=-(\mu^2+\nu_B^2)J_1(s\mu ).
	\end{align}
	In particular, this allows us to write
	\begin{align}
		\int_{0}^{\infty} s\left( 3\left(\beta+\nu_B\delta\right) -\bar{\mathcal{L}}\left[\gamma\right] \right)J_1(s\mu)\,\text{d}s =  \left(U-\dot{R}_B\right)\frac{4\mu}{\left(\mu^2+\nu_B^2\right)^{1/2}},
	\end{align}
	where the second equality follows from two integrations by parts if $\delta$ is bounded at $0$ and $\infty$. This allows us to rewrite Eq.~\eqref{eq:hydintermediatestep} as
	\begin{align}
		\int_{0}^{\infty}s\left(  3\left(\beta+\nu_B\delta\right) -\bar{\mathcal{L}}\left[\gamma\right] \right) J_1(s\mu)\,\text{d}s = \left(U-\dot{R}_B\right)\frac{4\mu}{\left(\mu^2+\nu_B^2\right)^{1/2}},
	\end{align}
	or, after applying another Hankel transform,
	\begin{align}
		3\left(\beta+\nu_B\delta\right) -\bar{\mathcal{L}}\left[\gamma\right]= 4(U-\dot{R}_B) s^{-2}\left(1+\nu_B s\right)e^{-\nu_B s}.
	\end{align}
	The other three boundary conditions Eqs.~\eqref{eq:BChyd1}-\eqref{eq:BChyd3} (no slip on the colloid and no penetration on the bubble) yield three linear equations which can be used to express $\alpha$, $\beta$ and $\delta$ in terms of $\gamma$. Then, we arrive at the modified Bessel equation,
	\begin{align}\label{eq:ODEhyd}
		\gamma''(s)+\frac{\gamma'(s)}{s}-\left(\frac{1}{s^2}+\nu_B^2\right)\gamma(s) = (U-\dot{R}_B)\frac{2\left(1+\nu_Bs\right)e^{-\nu_Bs}}{s^2}.
	\end{align}
	Considering the behaviour for small and large $s$, we find that the appropriate boundary conditions for a regular solution are $\gamma(0)=-2(U-\dot{R}_B)$ and $\gamma \to 0$ as $s\to\infty$, which is consistent with the integration by parts performed earlier. Since $\gamma$ is clearly linear in both $U$ and $\dot{R}_B$, a complete solution to the hydrodynamic problem for any bubble size $\nu_B$ may be found by solving Eq.~\eqref{eq:ODEhyd}. In fact, it is possible to do so by inspection. The result is
	\begin{align}
		\gamma(s)=-2(U-\dot{R}_B)e^{-\nu_Bs}.
	\end{align}
	Bootstrapping our way up again to find the coefficient $B$ we find the values
	\begin{align}
		B_\text{mot}=2U\times & \bigg[-2 \left((4 \nu_B+2) s^2+2 (\nu_B+1) s+1\right) w^{\nu_B+1}+\left(2 s^2-2 s+1\right) w^{2 \nu_B+1}\nonumber\\
		&+(1-2 \nu_B s) w^{\nu_B+2}+(2 \nu_B
		s+1) w^{\nu_B}-2 w^{2 \nu_B+2}+\left(2 s^2+2 s+1\right) w\bigg]\bigg{/}\nonumber\\
		&\big[s \left(1-4 (\nu_B+1) s w^{\nu_B+1}-w^{2 \nu_B+2}\right)\big],\\
		B_\text{gro}=-\dot{R}_B\times&  \bigg[2 \nu_B^2 s^2 (2 \nu_B s+1) w^{\nu_B}+2 \left\{2 \nu_B^2 (2 \nu_B+3) s^4+2 \nu_B \left(\nu_B^2+3 \nu_B+3\right) s^3\right.\nonumber\\
		&\left.+3 \left(\nu_B^2+4
		\nu_B+2\right) s^2+6 (\nu_B+1) s+3\right\} w^{\nu_B+1}-\left\{4 \nu_B^2 \left(8 \nu_B^2+12 \nu_B+3\right) s^4\right.\nonumber\\
		&\left.+4 \nu_B \left(6 \nu_B^2+11
		\nu_B+3\right) s^3+6 \left(3 \nu_B^2+4 \nu_B+1\right) s^2+6 (2 \nu_B+1) s+3\right\}w^{2 \nu_B+1}\nonumber\\
		&+\left(-8 \nu_B^3 s^3-6 \nu_B^2 s^2+6\right) w^{2\nu_B+2}-\left(8\nu_B^3 s^3+8 \nu_B^2 s^2+6 \nu_B s+3\right) w^{3 \nu_B+2}\nonumber\\
		&+(6 \nu_B s-3) w^{\nu_B+2}-3 \left(2 s^2+2 s+1\right) w\bigg]\bigg{/}\nonumber\\
		&\big[\nu_B^2 s^3 \left(1-4 (\nu_B+1) s w^{\nu_B+1}-w^{2 \nu_B+2}\right)\big],
	\end{align}
	where $w=e^{-2s}$. For efficient numerical evaluation, we recommend that asymptotic expressions should be used in form of a Taylor expansion in $s$ for $s\lesssim\min(0.1,0.1/\nu_B)$ and the leading order in $w$, $\mathcal{O}\left(w^{\min(1,\nu_B)}\right)$, for $s\gtrsim 10$. As $s\to 0$, the coefficient $A_\text{gro}\sim s^{-2}\to\infty$ diverges, which according to Eqs.~\eqref{eq:forceintegral1}-\eqref{eq:forceintegral2} leads to an infinite repulsive force between the two spheres. The origin and regularisation of this force is discussed in the next section.
	
		\subsection{Discussion of the pressure in the gap between bubble and colloid}\label{sec:pressure}
		In this paper we consider a setup in which the bubble and the colloid touch at a single point. As discussed in the main text, solving the hydrodynamic problem leads to infinite forces acting on the individual spheres, even though the net force on the system is finite. Since this is unrealistic, it is necessary to justify the physical validity of our results. In this section we investigate this divergence by solving for the flow near the contact point using a local asymptotic expansion of the Stokes equations. We demonstrate that the pressure diverges quadratically with a sign depending on the boundary condition, and hence that the divergence of the force due to having a singular contact point is logarithmic. As such, it is easily regularised even for small contact areas or separation between the spheres.
		
		For clarity, we stick to dimensional units in this section. Working in cylindrical coordinates $\{\rho,z\}$, the surface of the bubble is given by $z=h^+=R_B(1-\cos\theta^+)$ where $\tan\theta^+=\rho/R_B$. Expanding the cosine for small $\rho$ we arrive at the leading order expression $h^+=\rho^2/2R_B$. Similarly, we can expand the bubble surface as $z=-h^-=- \rho^2/2R_c$. Since the correct asymptotic expansion of the Stokes equation depends on the shortest length scale, we define $r_0=\min(R_b(t),R_c)$ and $\varepsilon=\rho/r_0$, which is our asymptotic expansion parameter. Geometrically we then have that at leading order
		\begin{align}
			h^+=\frac{r_0^2}{2R_B}\varepsilon^2,\quad 	h^-=\frac{r_0^2}{2R_c}\varepsilon^2.
		\end{align}
		The length scales are given by $\rho\sim r_0\varepsilon$ and $z\sim r_0\varepsilon^2$. Since $\varepsilon\ll 1$ by assumption, we have $\partial_z \gg \partial_\rho$ as in lubrication theory. By inspection, the leading-order solution for velocity and pressure is given by
		\begin{align}
			u_\rho\equiv u &= \varepsilon u_1+ \mathcal{O}(\varepsilon^2)\\
			u_z\equiv w &= \varepsilon^2 w_2 + \mathcal{O}(\varepsilon^3)\\
			p &= \varepsilon^{-2} p_{-2} + \mathcal{O}(\varepsilon^{-1}).
		\end{align}
		Introducing the scaled variable
		\begin{align}
			 \bar{z}\equiv\frac{2z}{r_0\varepsilon^2}=\frac{r_0}{R_B}\frac{z}{h^+}=\frac{r_0}{R_c}\frac{z}{h^-},
		\end{align}
		we find that the leading order terms in the incompressible Stokes equations become
		\begin{align}
			\partial_{\bar{z}\bar{z}}u_1 &= -\frac{r_0}{2\mu}p_{-2},\\
			u_1+\partial_{\bar{z}}w_2 &=0.
		\end{align}
		Additionally we have a continuity condition of the deforming bubble interface due to growth,
		\begin{align}
			\frac{Dh^+}{Dt}\equiv \partial_t h + u_\rho\vert_{z=h^+} \times\partial_\rho h &= u_z\vert_{z=h^+}\nonumber\\
			\Rightarrow -\frac{r_0^2 \dot{R}_B}{2R_B^2}+u_1\vert_{\bar{z}=r_0/R_B}\times\frac{r_0}{R_B}&=w_2\vert_{\bar{z}=r_0/R_B}.
		\end{align}
		
		For a \textbf{rigid} bubble, the boundary conditions to this system are $u_1=w_2=0$ at $\bar{z}=-r_0/R_c$ and $\bm{u}=-\dot{R}_B(\bm{\nu}-\bm{z})$ at $z=h^+$, where $\bm{\nu}=\cos\theta^+ \bm{z}-\sin\theta^+\bm{\rho}$. At leading order this second condition becomes
		\begin{align}
			u_1 = \dot{R}_B\frac{r_0}{R_B},\quad w_2 = \frac{\dot{R}_B}{2}\frac{r_0^2}{R_B^2},\quad\text{at }\bar{z}=r_0/R_B.
		\end{align}
		It may be checked that this is compatible with the continuity condition, and we note that $Dh/Dt$ is positive, so material points on the bubble move away from the colloid as the bubble expands. In particular, $\partial_t{h}$ and $u_z\vert_{z=h}$ have opposite signs.  Solving the Stokes equations with these boundary conditions leads to messy expressions for $u_1(\bar{z})$ and $w_2(\bar{z})$, as well as the dominant pressure term,
		\begin{align}
			p_{-2} = -12\mu\dot{R}_B\frac{R_BR_c^2\left(R_B+2R_c\right)}{r_0^2\left(R_B+R_c\right)^3}.
		\end{align}
	 	Hence $p\sim\rho^{-2}$ as $\rho\to 0$, so the area integrated stress exerted by the spheres diverges logarithmically. Since the pressure is strictly negative, it corresponds to an \textbf{\emph{attraction}} of the two spheres. Intuitively, this may be explained by visualising a stretching of the bubble surface as that of a balloon. While for fixed $\rho$ the gap width $h$ reduces as the bubble expands, individual points on the bubble surface move radially outwards at $\mathcal{O}(\varepsilon)$, and upwards at $\mathcal{O}(\varepsilon^2)$. Because of incompressibility fluid is sucked into the gap to compensate for the entrainment by the expanding bubble surface, creating an attraction.
		
		This is not the case if the bubble surface exerts \textbf{no shear stress}. In this case the upper boundary conditions are replaced by $\partial_{\bar{z}}u_1=0$ at $\bar{z}=r_0/R_B$ and the continuity condition. This time the solution for the pressure is 
		\begin{align}
			p_{-2} = 6\mu\dot{R}_B\frac{R_BR_c^3}{r_0^2\left(R_B+R_c\right)^2\left(2R_B+5R_c\right)},
		\end{align}
		which is strictly positive. Therefore, the bubble and colloid \textbf{\emph{repel}} each other with this boundary condition, as one might expect intuitively. Differently from the rigid case, we have 
		\begin{equation}
		\frac{Dh}{Dt}=-\frac{r_0^2 (R_b + R_c)}{R_b^2 (2 R_b + 5 R_c)}<0,
		\end{equation}
		 so material points on the bubble move in to close the gap as the bubble expands. As before, the divergence of the pressure is quadratic, which means the repulsive force is easily regularised.
	
	\section{Discussion of possible physics leading to premature bubble disappearance}\label{sec:bubbleBgone}
	
	\begin{figure}[t!]
		\centering
		\includegraphics[width=0.4\textwidth]{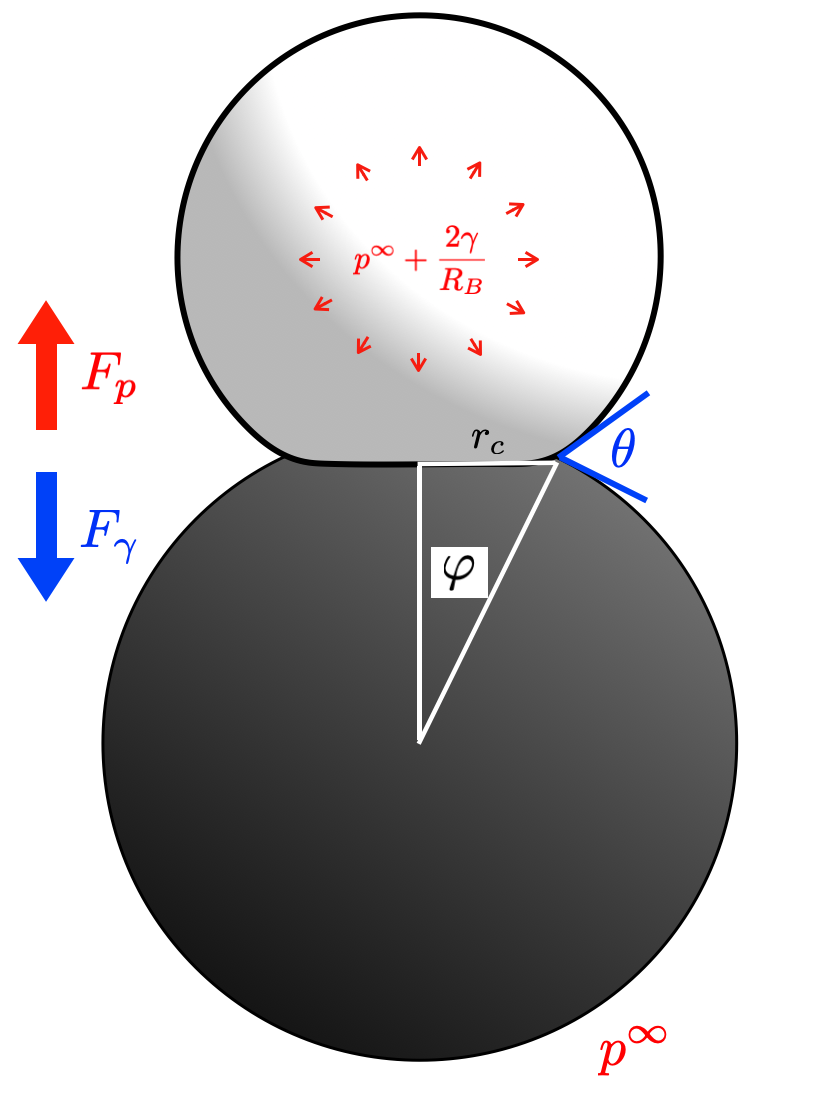}
		\caption{Sketch of a more realistic geometry of bubble growth in which the bubble forms a spherical cap on the colloid. The contact angle $\theta$ and coverage angle $\phi$ are exaggerated for illustrative purposes. An internal pressure detachment force $F_p$ and a contact line adhesion force $F_\gamma$ compete in attaching the bubble to the colloid.}\label{fig:cap}
	\end{figure}
	
	As outlined in section \ref{sec:comp_exp}, our model predicts the slow diffusive growth of the bubble to a final size that, in large regions of the parameter space, is significantly larger than the colloid. In experiments, these bubbles are generically not observed~\cite{manjare2012bubble,manjare2013bubble}, which led us to question possible mechanisms of premature bubble detachment. To this end it is necessary to revisit an assumption we have made in developing this model, namely that the bubble and colloid are both spherical and touch at an infinitesimally small point. \revision{If there is partial wetting, the bubble may form a spherical cap on the colloid with a small but non-zero contact radius that is determined by the surface properties of the colloid and the coefficients of surface tension between the gas, liquid, and solid materials that meet there.} This is illustrated in Fig.~\ref{fig:cap}. Here we introduce two angles, the contact angle $\theta$, measured in the liquid phase, and the coverage angle $\phi$ which is related to the radius of the contact line $r_c=\sin\phi$ in scaled units.
	
	As a first check, we asked whether buoyancy would be sufficient to detach the bubble from the surface. For this, we compared the scaling of adhesive force along the contact line against buoyancy on the bubble volume as a function of the contact radius $r_c$. The point at which they balance is known as the Fritz radius $r_f$~\cite{rodriguez2015generation}, and the bubble detaches only if the true contact is smaller. From the scales of our problem we can estimate the Fritz radius as
	\begin{align}
		r_f \sim \frac{\tfrac{4}{3}\pi \rho g L^3}{2\pi\gamma} \sim \SI{0.1}{\nano\metre},
	\end{align}
	where we assumed a length $L=\mathcal{O}(\SI{10}{\micro\meter})$. Despite the $L^3$-scaling rapidly increasing this radius for larger bubbles, this is so small that the bubble needs to cover only a vanishing fraction of the colloid surface in order to be tethered. In conclusion it seems unlikely that buoyancy is the cause of premature detachment. 
	
	A second check concerns the internal pressure in the bubble pushing against the colloid, $F_p$, compared to the contact line adhesive force from surface tension $F_\gamma$. Since the pressure difference between the bubble and the surrounding fluid is dominated by capillary pressure, we can approximate these as
	\begin{align}
		F_p &= \iint_\text{contact area} \frac{2\gamma}{R_b}\bm{n}\cdot\hat{\bm{z}}\,\text{d}A \nonumber\\
		&= \frac{2\gamma}{R_B}\pi\sin^2\phi,\\
		F_\gamma &=\int_\text{contact line} \gamma \bm{t}\cdot\hat{\bm{z}}\,\text{d}S\nonumber\\
		&= 2\pi\gamma\sin\phi\sin\left(\theta-\phi\right),
	\end{align}
	where $\bm{n}$ is the normal to the colloid and $\bm{t}$ is tangential to the bubble. We ask whether the relative strength of these two forces depends on the angles involved, and whether this leads to a condition on the bubble radius $R_B$ to which the angles $\theta$ and $\phi$ are geometrically related. It turns out however the reverse is true, and these forces always balance. Specifically, by considering the triangle formed by the sphere centres and the contact point it follows from the law of sines that
	\begin{align}
		R_B = \frac{\sin\phi}{\sin(\theta-\phi)},
	\end{align}
	and the result $F_p=F_\gamma$ follows from an easy substitution. While this may seem surprising, it just means that the contact angle $\theta$ is not controlled dynamically by the state of the system, but by the surface and material properties, which are unknown. \revision{They would be even more important when a moving contact line, and possibly pinning or surface roughness played a role.} We also made one implicit assumption in this argument, which is that the fluid pressure is uniform. As shown in section \ref{sec:pressure}, this is not the case in the gap, which could tip the balance in this argument. However, since the sign of this gap pressure is not dynamic and the bubble does not detach immediately, we suspect that the true boundary condition is approximately no slip, and the gap pressure indeed acts to further glue the bubble to the colloid.
	
	In conclusion, we could not find an argument based on the intrinsic physics of the growth process that supports premature detachment. However, in separate experiments involving bubble formation in a thin film with immersed catalytic colloids, it was observed that coalescence with an external air-water interface can lead to the sudden disappearance of the bubble~\cite{yang2017peculiar}. While the depth of this film was not reported in previous work~\cite{manjare2012bubble}, we strongly suspect that the same external limitation of the system was causing the bubble disappearance there, rather than sudden inertial collapse that is hard to justify in an otherwise entirely non-inertial system. Arguing along similar lines, hydrodynamic interactions with the rigid interface on which the colloids are located may also hinder bubble growth. A detailed numerical investigation of this confinement however lies beyond the scope of this study. 
	
			\bibliographystyle{vancouver}
\bibliography{bibliography_bubble_paper}

\begin{thebibliography}{10}

\bibitem{paxton04}
Paxton WF, Kistler KC, Olmeda CC, Sen A, St~Angelo SK, Cao Y, et~al.
\newblock Catalytic nanomotors: autonomous movement of striped nanorods.
\newblock Journal of the American Chemical Society. 2004;126(41):13424--13431.

\bibitem{elgeti2015physics}
Elgeti J, Winkler RG, Gompper G.
\newblock Physics of microswimmers - single particle motion and collective
  behavior: a review.
\newblock Reports on progress in physics. 2015;78(5):056601.

\bibitem{illien2017fuelled}
Illien P, Golestanian R, Sen A.
\newblock `Fuelled' motion: phoretic motility and collective behaviour of
  active colloids.
\newblock Chemical Society Reviews. 2017;46(18):5508--5518.

\bibitem{moran2017phoretic}
Moran JL, Posner JD.
\newblock Phoretic self-propulsion.
\newblock Annual Review of Fluid Mechanics. 2017;49:511--540.

\bibitem{nelson2010microrobots}
Nelson BJ, Kaliakatsos IK, Abbott JJ.
\newblock Microrobots for minimally invasive medicine.
\newblock Annual review of biomedical engineering. 2010;12:55--85.

\bibitem{jeon2019magnetically}
Jeon S, Kim S, Ha S, Lee S, Kim E, Kim SY, et~al.
\newblock Magnetically actuated microrobots as a platform for stem cell
  transplantation.
\newblock Science Robotics. 2019;4(30):eaav4317.

\bibitem{golestanian2007designing}
Golestanian R, Liverpool T, Ajdari A.
\newblock Designing phoretic micro-and nano-swimmers.
\newblock New Journal of Physics. 2007;9(5):126.

\bibitem{chamolly2019stochastic}
Chamolly A, Lauga E.
\newblock Stochastic dynamics of dissolving active particles.
\newblock The European Physical Journal E. 2019;42:1--15.

\bibitem{chamolly2020irreversible}
Chamolly A, Lauga E, Tottori S.
\newblock Irreversible hydrodynamic trapping by surface rollers.
\newblock Soft matter. 2020;16(10):2611--2620.

\bibitem{tottori2012magnetic}
Tottori S, Zhang L, Qiu F, Krawczyk KK, Franco-Obreg{\'o}n A, Nelson BJ.
\newblock Magnetic helical micromachines: fabrication, controlled swimming, and
  cargo transport.
\newblock Advanced materials. 2012;24(6):811--816.

\bibitem{walther2012soft}
Walther A, M{\"u}ller AH.
\newblock Soft, Nanoscale Janus Particles by Macromolecular Engineering and
  Molecular Self-assembly.
\newblock In: Janus Particle Synthesis, Self-Assembly and Applications. Royal
  Society of Chemistry; 2012. p. 1--28.

\bibitem{li2018development}
Li J, Li X, Luo T, Wang R, Liu C, Chen S, et~al.
\newblock Development of a magnetic microrobot for carrying and delivering
  targeted cells.
\newblock Science robotics. 2018;3(19):eaat8829.

\bibitem{soto2021smart}
Soto F, Karshalev E, Zhang F, Esteban Fernandez~de Avila B, Nourhani A, Wang J.
\newblock Smart materials for microrobots.
\newblock Chemical Reviews. 2021;122(5):5365--5403.

\bibitem{ghosh2009controlled}
Ghosh A, Fischer P.
\newblock Controlled propulsion of artificial magnetic nanostructured
  propellers.
\newblock Nano letters. 2009;9(6):2243--2245.

\bibitem{mou2015single}
Mou F, Li Y, Chen C, Li W, Yin Y, Ma H, et~al.
\newblock Single-Component TiO2 Tubular Microengines with Motion Controlled by
  Light-Induced Bubbles.
\newblock Small. 2015;11(21):2564--2570.

\bibitem{michelin2014phoretic}
Michelin S, Lauga E.
\newblock Phoretic self-propulsion at finite P{\'e}clet numbers.
\newblock Journal of fluid mechanics. 2014;747:572--604.

\bibitem{moran2011electrokinetic}
Moran JL, Posner JD.
\newblock Electrokinetic locomotion due to reaction-induced charge
  auto-electrophoresis.
\newblock Journal of Fluid Mechanics. 2011;680:31--66.

\bibitem{gibbs2009autonomously}
Gibbs JG, Zhao YP.
\newblock Autonomously motile catalytic nanomotors by bubble propulsion.
\newblock Applied Physics Letters. 2009;94(16):163104.

\bibitem{manjare2012bubble}
Manjare M, Yang B, Zhao YP.
\newblock Bubble driven quasioscillatory translational motion of catalytic
  micromotors.
\newblock Physical review letters. 2012;109(12):128305.

\bibitem{manjare2013bubble}
Manjare M, Yang B, Zhao YP.
\newblock Bubble-propelled microjets: Model and experiment.
\newblock The Journal of Physical Chemistry C. 2013;117(9):4657--4665.

\bibitem{gallino2018physics}
Gallino G, Gallaire F, Lauga E, Michelin S.
\newblock Physics of Bubble-Propelled Microrockets.
\newblock Advanced Functional Materials. 2018;28(25):1800686.

\bibitem{yang2001numerical}
Yang Z, Dinh TN, Nourgaliev R, Sehgal B.
\newblock Numerical investigation of bubble growth and detachment by the
  lattice-Boltzmann method.
\newblock International Journal of Heat and Mass Transfer. 2001;44(1):195--206.

\bibitem{zeng1993unified}
Zeng L, Klausner J, Mei R.
\newblock A unified model for the prediction of bubble detachment diameters in
  boiling systems--I. Pool boiling.
\newblock International Journal of Heat and Mass Transfer.
  1993;36(9):2261--2270.

\bibitem{fomin2013propulsion}
Fomin VM, Hippler M, Magdanz V, Soler L, Sanchez S, Schmidt OG.
\newblock Propulsion mechanism of catalytic microjet engines.
\newblock IEEE Transactions on Robotics. 2013;30(1):40--48.

\bibitem{li2014hydrodynamics}
Li L, Wang J, Li T, Song W, Zhang G.
\newblock Hydrodynamics and propulsion mechanism of self-propelled catalytic
  micromotors: Model and experiment.
\newblock Soft Matter. 2014;10(38):7511--7518.

\bibitem{wang2017viscosity}
Wang Z, Chi Q, Liu L, Liu Q, Bai T, Wang Q.
\newblock A viscosity-based model for bubble-propelled catalytic micromotors.
\newblock Micromachines. 2017;8(7):198.

\bibitem{wang2018dynamic}
Wang Z, Chi Q, Bai T, Wang Q, Liu L.
\newblock A dynamic model of drag force for catalytic micromotors based on
  navier--stokes equations.
\newblock Micromachines. 2018;9(9):459.

\bibitem{lohse2015surface}
Lohse D, Zhang X, et~al.
\newblock Surface nanobubbles and nanodroplets.
\newblock Reviews of modern physics. 2015;87(3):981.

\bibitem{rodriguez2015generation}
Rodr{\'\i}guez-Rodr{\'\i}guez J, Sevilla A, Mart{\'\i}nez-Baz{\'a}n C, Gordillo
  JM.
\newblock Generation of microbubbles with applications to industry and
  medicine.
\newblock Annual review of fluid mechanics. 2015;47:405--429.

\bibitem{happel2012low}
Happel J, Brenner H.
\newblock Low Reynolds number hydrodynamics: with special applications to
  particulate media. vol.~1.
\newblock Springer Science \& Business Media; 2012.

\bibitem{lauga2009hydrodynamics}
Lauga E, Powers TR.
\newblock The hydrodynamics of swimming microorganisms.
\newblock Reports on progress in physics. 2009;72(9):096601.

\bibitem{zhai2019precise}
Zhai W, Song Y, Gao Z, Fan JB, Wang S.
\newblock Precise synthesis of polymer particles spanning from anisotropic
  Janus particles to heterogeneous nanoporous particles.
\newblock Macromolecules. 2019;52(9):3237--3243.

\bibitem{kaewsaneha2013janus}
Kaewsaneha C, Tangboriboonrat P, Polpanich D, Eissa M, Elaissari A.
\newblock Janus colloidal particles: preparation, properties, and biomedical
  applications.
\newblock ACS applied materials \& interfaces. 2013;5(6):1857--1869.

\bibitem{sabass2012dynamics}
Sabass B, Seifert U.
\newblock Dynamics and efficiency of a self-propelled, diffusiophoretic
  swimmer.
\newblock The Journal of chemical physics. 2012;136(6).

\bibitem{sharifi2013diffusiophoretic}
Sharifi-Mood N, Koplik J, Maldarelli C.
\newblock Diffusiophoretic self-propulsion of colloids driven by a surface
  reaction: the sub-micron particle regime for exponential and van der Waals
  interactions.
\newblock Physics of Fluids. 2013;25(1).

\bibitem{inoue2022ostwald}
Inoue S, Kimura Y, Uematsu Y.
\newblock Ostwald ripening of aqueous microbubble solutions.
\newblock The Journal of Chemical Physics. 2022;157(24).

\bibitem{zhou2022solid}
Zhou J, Wei C, Dong Y.
\newblock How solid--liquid adsorption affects the liquid--vapor interface in
  pores.
\newblock Vadose Zone Journal. 2022;21(5):e20214.

\bibitem{plesset1977bubble}
Plesset MS, Prosperetti A.
\newblock Bubble dynamics and cavitation.
\newblock Annual review of fluid mechanics. 1977;9(1):145--185.

\bibitem{lide2004crc}
Lide DR.
\newblock CRC handbook of chemistry and physics. vol.~85.
\newblock CRC press; 2004.

\bibitem{yu2019porosity}
Yu W, Batchelor-McAuley C, Chang X, Young NP, Compton RG.
\newblock Porosity controls the catalytic activity of platinum nanoparticles.
\newblock Physical Chemistry Chemical Physics. 2019;21(36):20415--20421.

\bibitem{chang2020hydrogen}
Chang X, Batchelor-McAuley C, Compton RG.
\newblock Hydrogen peroxide reduction on single platinum nanoparticles.
\newblock Chemical Science. 2020;11(17):4416--4421.

\bibitem{yang2017peculiar}
Yang F, Manjare M, Zhao Y, Qiao R.
\newblock On the peculiar bubble formation, growth, and collapse behaviors in
  catalytic micro-motor systems.
\newblock Microfluidics and Nanofluidics. 2017;21(1):6.

\bibitem{chamolly2020stokes}
Chamolly A, Lauga E.
\newblock Stokes flow due to point torques and sources in a spherical geometry.
\newblock Physical Review Fluids. 2020;5(7):074202.

\bibitem{strogatz2018nonlinear}
Strogatz SH.
\newblock Nonlinear dynamics and chaos with student solutions manual: With
  applications to physics, biology, chemistry, and engineering.
\newblock CRC press; 2018.

\bibitem{gao2012hydrogen}
Gao W, Uygun A, Wang J.
\newblock Hydrogen-bubble-propelled zinc-based microrockets in strongly acidic
  media.
\newblock Journal of the American Chemical Society. 2012;134(2):897--900.

\bibitem{cooley1969slow}
Cooley M, O'Neill M.
\newblock On the slow motion of two spheres in contact along their line of
  centres through a viscous fluid.
\newblock In: Mathematical Proceedings of the Cambridge Philosophical Society.
  vol.~66. Cambridge University Press; 1969. p. 407--415.

\bibitem{reed1974slow}
Reed L, Morrison~Jr F.
\newblock The slow motion of two touching fluid spheres along their line of
  centers.
\newblock International Journal of Multiphase Flow. 1974;1(4):573--584.

\bibitem{michelin2018collective}
Michelin S, Gu{\'e}rin E, Lauga E.
\newblock Collective dissolution of microbubbles.
\newblock Physical Review Fluids. 2018;3(4):043601.

\end{thebibliography}
\end{document}